\documentclass[aps]{revtex4}
\usepackage{amsfonts}
\usepackage{amsmath}
\usepackage{amssymb,epsf}
\usepackage{graphicx}
\usepackage{color}

\begin{document}

\title{A new approach toward geometrical concept of black hole thermodynamics}
\author{Seyed Hossein Hendi$^{1,2}$\footnote{email address: hendi@shirazu.ac.ir}, Shahram Panahiyan$^{1}$\footnote{
email address: ziexify@gmail.com}, Behzad Eslam
Panah$^{1}$\footnote{email address:
behzad$_{-}$eslampanah@yahoo.com} and Mehrab
Momennia$^{1}$\footnote{email address: momennia1988@gmail.com}}
\affiliation{$^1$ Physics Department and Biruni Observatory,
College of Sciences, Shiraz University, Shiraz 71454, Iran\\
$^2$ Research Institute for Astronomy and Astrophysics of Maragha
(RIAAM), Maragha, Iran}

\begin{abstract}
Motivated by the energy representation of Riemannian metric, in
this paper we study different approaches toward the geometrical
concept of black hole thermodynamics. We investigate
thermodynamical Ricci scalar of Weinhold, Ruppeiner and Quevedo
metrics and show that their number and location of divergences do
not coincide with phase transition points arisen from heat
capacity. Next, we introduce a new metric to solve these problems.
We show that the denominator of the Ricci scalar of the new metric
contains terms which coincide with different types of phase
transitions. We elaborate the effectiveness of the new metric and
shortcomings of the previous metrics with some examples.
Furthermore, we find a characteristic behavior of the new
thermodynamical Ricci scalar which enables one to distinguish two
types of phase transitions. In addition, we generalize the new
metric for the cases of more than two extensive parameters and
show that in these cases the divergencies of thermodynamical Ricci
scalar coincide with phase transition points of the heat capacity.
\end{abstract}

\maketitle

\section{Introduction}

One of the succinct semi-classical approaches for investigating
the quantum nature of gravity is via black hole thermodynamics in
AdS spacetime which is dual to that of a field theory in one
dimension fewer \cite{ADSCFT}. Besides, an interesting development
in the black holes studies comes from the fact that black holes
are akin to thermodynamical system which can be essentially
described by the laws of thermodynamics re-expressed in terms of
properties of black holes
\cite{Davies,Hawking,Bekenstein,Bardeen,Hut}.

Scientific researches in the black hole physics have presented a very deep
and fundamental relationship between quantum gravity, holography and
thermodynamics. Among these researches, the special interest is attracted by
AdS black holes because some of them lead to the very interesting phenomenon
called Hawking-Page phase transition \cite{HP}. Phase transition plays an
important role in order to explore thermodynamical properties of a system
near the critical point. Usually, phase transitions are denoted by a
discontinuity of a state space variable, specially heat capacity \cite{HC}.

Motivated by large applications of the geometrical concept of
thermodynamics in the black hole phase transition, in this paper,
we investigate strengths and shortcomings of different approaches
toward the matter. First attempt was done by Weinhold
\cite{Weinhold} which introduced a metric on the space of
equilibrium states where its components are given as the Hessian
of the internal energy. Then, Ruppeiner introduced a metric which
is defined as the negative Hessian of entropy with respect to the
internal energy and other extensive quantities of a
thermodynamical system \cite{Ruppeiner}. It was shown that these
two metrics are conformally equivalent to each other where the
temperature is the conformal factor \cite{Salamon}. Recently,
Quevedo \cite{Quevedo} proposed a Legendre invariant metric, in
which solved some of problems in Weinhold/Ruppeiner methods.

The basic motivation for considering the geometrothermodynamics
comes from the fact that this formalism describes in an invariant
way the thermodynamic properties of a given thermodynamical system
in terms of geometric structures. Although extracting phase
transition in terms of curvature singularity is the main reason to
consider the geometrical approach in thermodynamics, there are
several examples in which the curvature singularities of the known
metrics (Weinhold and Ruppeiner metrics and their Legendre
transformations) are not located at the phase transition points
and they have number of singularities in which are before/after
the phase transition points \cite{Shen,Aman}.

In this paper, in order to solve the mentioned problems, we introduce a
metric where its curvature singularities are, exactly, located at the phase
transition points. As we see later, this effective metric has different
structure from the other known metrics. For this claim, we consider the
black hole solutions with Born-Infeld (BI) and Maxwell sources in three and
four dimensions and investigate phase transitions in various
geometrothermodynamical methods.


\section{Heat Capacity}

In order to investigate the local stability of a black hole, one
can use various methods related to different ensembles. In
principle thermal stability can be carried out in the grand
canonical ensemble by finding the determinant of the Hessian
matrix of $M(X_{i})$ with respect to its extensive variables
$X_{i}$ \cite{Cvetic,Caldarelli}. For static charged black holes,
we usually regard the mass $M$ as a function of the entropy $S$
and the charge $Q$. The number of thermodynamic variables depends
on the ensemble that is used. The most well known approach for
studying phase transition is in the context of canonical ensemble,
hence the heat capacity of systems. Depending on the number of
extensive parameters, the behavior of Hessian matrix is highly
sensitive \cite{Sens} and therefore most of physicists prefer to
work in canonical ensemble. In this ensemble, the positivity of
the heat capacity is sufficient to ensure thermal stability. In
addition the system is considered to be in fixed charged and the
heat capacity has the following form
\begin{equation}
C_{Q}=\frac{M_{S}}{M_{SS}},  \label{heat}
\end{equation}%
where $M_{S}=\left( \frac{\partial M}{\partial S}\right) _{Q}$ and
$M_{SS}=\left( \frac{\partial ^{2}M}{\partial S^{2}}\right) _{Q}$
are regular functions. Now we regard two different types of phase
transitions. In type one, the changes in signature of the heat
capacity is representing a phase transition. In other words, the
roots of the heat capacity in this case are representing phase
transition points which means one should solve $M_{S}=0$. Phase
transition concerning to the divergency of the heat capacity is
denoted by type two. It means the singular points of the heat
capacity are representing the phase transitions. Thus, we should
consider $M_{SS}=0$ to obtain the phase transition of type two.

In order to have a fitting geometrical approach for studying phase
transitions, the thermodynamical Ricci scalar (TRS) must diverges in the
mentioned both types of phase transitions. In what follows, we present a
brief review study of several geometrical approaches with their
shortcomings, and then, we propose a new effective metric concerning this
issue.


\section{Weinhold metric}

In Weinhold method, one is considering appropriate extensive parameters such
as entropy, electric charge and angular momentum, and their related
intensive quantities such as temperature, electric potential and angular
velocity with the mass of the black holes as a potential. The Weinhold
metric is given by \cite{Weinhold}
\begin{equation}
ds_{W}^{2}=Mg_{ab}^{W}dX^{a}dX^{b},  \label{Wein}
\end{equation}%
where $g_{ab}^{W}=\partial ^{2}M\left( X^{c}\right) /\partial
X^{a}\partial X^{b}$ and also $X^{a}\equiv X^{a}\left(
S,N^{i}\right) $, where $N^{i}$ denotes other extensive variables
of the system. Considering a static charged black hole, one can
find the following expression for the denominator of Weinhold
Ricci scalar
\begin{equation}
denom(\mathcal{R}_{W})=\left( M_{SS}M_{QQ}-M_{SQ}^{2}\right) ^{2}M^{2}\left(
S,Q\right) ,  \label{denWein}
\end{equation}
where $M_{QQ}=\left( \frac{\partial ^{2}M }{\partial Q^{2}}
\right)_{S} $ and $M_{SQ}=\frac{\partial ^{2}M}{\partial S\partial
Q} $ and for consistency (with respect to the heat capacity
results), the roots of the Eq. (\ref{denWein}) should coincide
with two types of the mentioned phase transitions in the heat
capacity. It is easy to find that due to the structure
of the Eq. (\ref{denWein}), only in special case $M_{SQ}=0$ and nonzero $%
M_{QQ}$, the divergence points of the heat capacity coincide with
divergencies of the Weinhold Ricci scalar. In order to obtain consistent
results for the type one phase transition, the following fine tuning must be
hold
\begin{equation}
M_{SS}M_{QQ}-M_{SQ}^2=M_{S}. \label{FTW}
\end{equation}

Regarding various case studies and calculating $M$ and its
derivatives, we should note that Eq. (\ref{FTW}) is not always
satisfied. In addition, it is evident that for the case of
$M_{SS}=\frac{M_{SQ}^{2}}{M_{QQ}}$, there will be extra
divergencies for $\mathcal{R}_{W}$ which are not related to any
phase transition of the heat capacity. Therefore, the structure of
this part of denominator is in a way that may present extra
divergencies that do not coincide with any type of phase
transition points of the heat capacity.


\section{Ruppeiner metric}

The Ruppeiner metric is defined as \cite{Ruppeiner}
\begin{equation}
ds_{R}^{2}=g_{ab}^{R}dY^{a}dY^{b},  \label{Rupp}
\end{equation}
where $g_{ab}^{R}=-\partial ^{2}S\left( Y^{c}\right) /\partial Y^{a}\partial
Y^{b}$ and $Y^{a}\equiv Y^{a}\left( M,N^{i}\right) $. It was proved that
Weinhold and Ruppeiner metrics are related to each other by a Legendre
transformation \cite{Quevedo}
\begin{equation}
ds_{R}^{2}=-M T^{-1} g_{ab}^{W}dX^{a}dX^{b}.  \label{Rupp2}
\end{equation}

Applying this transformation, one can find the following relation for the
denominator of the Ricci scalar for this case
\begin{equation}
denom(\mathcal{R}_{R})=\left( M_{SS}M_{QQ}-M_{SQ}^{2}\right) ^{2}T(S,Q)
M^{2}\left( S,Q\right) .  \label{denRupp}
\end{equation}

This equation shows that the phase transitions of the heat
capacity coincide with the singularities of Ruppeiner Ricci scalar
only when $M_{SQ}^{2}=0$ and nonzero $M_{QQ}$. In addition,
regarding type one phase transitions, one finds due to existence
of the $T(S,Q)$ and the fact that $T(S,Q)=M_{S}$, the roots of the
heat capacity and some of the divergence points of Ruppeiner Ricci
scalar coincide. On the other hand, there may be some divergence
points of $\mathcal{R}_{R}$ that do not coincide with phase
transition points. These extra phase transition points are
originated from the zeroes of $\left(
M_{SS}M_{QQ}-M_{SQ}^{2}\right)$. As we mentioned before, in the
case of $M_{SQ}^{2}=0$, type two phase transitions of the heat capacity ($%
M_{SS}=0$) are covered by divergencies of the Ricci scalar of the
Ruppeiner metric. But in this case, we encounter with extra
divergencies related to the roots of $M_{QQ}=0$ which are not
related to any phase transition of the heat capacity. In addition,
in the case of nonzero $M_{SQ}^{2}$, the possible real roots of
$M_{QQ}=\frac{M_{SQ}^{2}}{M_{SS}}$ lead to the same extra
divergencies, which were observed in Weinhold metric.


\section{Quevedo metrics}

In order to remove the failures of the Weinhold and Ruppeiner metrics,
Quevedo proposed a new thermodynamical metric. The Quevedo metric can be
written as \cite{Quevedo}
\begin{equation}
ds_{Q}^{2}=\Omega \left( -M_{SS}dS^{2}+M_{QQ}dQ^{2}\right) ,  \label{Quev}
\end{equation}
where the (conformal) function $\Omega$ has one of the following
forms
\begin{equation}
\Omega =\left\{
\begin{array}{cc}
SM_{S}+QM_{Q}, & \text{case I} \\
SM_{S}, & \text{case II}%
\end{array}%
\right. .  \label{gabQ}
\end{equation}

In order to obtain the curvature singularity of the Quevedo metric, we
calculate the Ricci scalar. Although calculation of the Ricci scalar is
straightforward, analytically calculated results are too large. So, for the
sake of brevity we do not write the long equations of the Ricci scalar;
instead, we can use numerical analysis and some plots to investigate the
Ricci scalar's behavior. In addition, since we are looking for the
divergence points of the Ricci scalars, we can study the denominator of the
Ricci scalars with the following explicit forms
\begin{equation}
denom(\mathcal{R}_{Q1})=\left( SM_{S}+QM_{Q}\right) ^{3}M_{SS}^{2}M_{QQ}^{2},
\label{RQ1}
\end{equation}%
\begin{equation}
denom(\mathcal{R}_{Q2})=S^{3}M_{S}^{3}M_{SS}^{2}M_{QQ}^{2}.  \label{RQ2}
\end{equation}

Due to $M_{SS}$ being in the denominator of both Quevedo Ricci
scalars, the divergencies of the heat capacity and Quevedo Ricci
scalars coincide. Regarding type one phase transition, although
the roots of the heat capacity and divergencies of the Quevedo
Ricci scalar of case II coincide, for case I this coincidence
takes place only for vanishing $M_{Q}$ (which is in general a
nonzero function). It is worthwhile to mention the fact that there
exists an additional function $M_{QQ}^{2}$, and its roots provides
extra singular points for both cases of Quevedo Ricci scalars.
Although $M_{Q}$ and $M_{S}$ are independent from each other,
generally, for a nonzero $M_{Q}$ in the case I, it may be possible
to set $M_{S}=-\frac{Q}{S}M_{Q}$ for special choices of free
parameters. In this situation, one finds another divergence point
which may not coincide with any phase transition points of the
heat capacity.


\section{new metric}

In order to avoid extra singular points in TRS which do not coincide with
phase transitions of any type, and also to ensure all divergencies of the
TRS coincide with phase transition points of the both types, we introduce
the following new thermodynamical metric
\begin{equation}
ds_{new}^{2}=S\frac{M_{S}}{M_{QQ}^{3}}\left(
-M_{SS}dS^{2}+M_{QQ}dQ^{2}\right) .  \label{newmetric}
\end{equation}

It is worthwhile to mention that the new thermodynamical metric is defined
the same as the Quevedo metric with different conformal function. In new
metric, we have considered the total mass as thermodynamical potential with
entropy and electric charge as extensive parameters. In order to find the
geometrical behavior of new thermodynamical metric, we calculate the Ricci
scalar. It is a straightforward calculation to show that the numerator and
the denominator of TRS for this new metric is, respectively,
\begin{eqnarray}
num(\mathcal{R})
&=&6S^{2}M_{S}^{2}M_{QQ}M_{SS}^{2}M_{QQQQ}-6SM_{S}^{2}M_{QQ}^{2}M_{SS}M_{SSQQ}+2SM_{SQQ}^{2}M_{S}^{2}M_{QQ}M_{SS}
\notag \\
&&+2\left[SM_{S}M_{SSS}-\frac{1}{2}M_{SS}\left(SM_{SS}-M_{S}\right) \right]%
SM_{QQ}^{2}M_{S}M_{SQQ}-9S^{2}M_{QQQ}^{2}M_{S}^{2}M_{SS}^{2}  \notag \\
&&+4\left[ \frac{1}{4}M_{SQ}M_{SS}+M_{S}M_{SSQ}\right]
S^{2}M_{QQ}M_{S}M_{QQQ}+\left[
S^{2}M_{S}^{2}M_{SSQ}-S^{2}M_{SQ}M_{SS}M_{S}M_{SSQ}\right.  \notag \\
&&\left. SM_{QQ}M_{S}\left( SM_{SS}-M_{S}\right) M_{SS}-2\left(
S^{2}M_{SS}^{3}+M_{S}^{2}M_{SS}\right)
M_{QQ}+2S^{2}M_{SQ}^{2}M_{SS}^{2} \right] M_{QQ}^{2},
\label{NumNew}
\end{eqnarray}
and
\begin{equation}
denom(\mathcal{R})=S^{3}M_{S}^{3}M_{SS}^{2}.  \label{denNew}
\end{equation}

Regarding a typical black hole with nonzero horizon radius,
$r_{+}$ (and also nonzero entropy), one can find that, in general,
the finite mass is an analytic smooth function of its variables.
Smooth function is often used technically to mean a function that
has derivatives of all orders everywhere in its domain. Therefore,
one may take into account that $M$ has regular derivatives of all
orders with respect to the extensive parameters \cite{Singular}.

Taking into account the regular numerator of TRS in Eq.
(\ref{NumNew}) with the denominator of TRS, Eq. (\ref{denNew}), we
ensure that all the phase transition points of the type one and
two coincide with divergencies of the mentioned TRS and there is
no extra term that may provide extra divergencies.

We should note that, in general, the derivatives of all orders of
$M$ (such as $M_{QQ}$, $M_{SS}$, $M_{SQ}$, $M_{SSQ}$ and so on)
are independent from each other. In addition, we should mention
that for the case of vanishing $M_{S}$ or $M_{SS}$, the numerator
of TRS has nonzero value, but denominator of TRS vanishes. In the
case of $M_{S}=M_{SS}=0$, although it is clear that both numerator
and denominator vanish, the denominator approaches zero faster
than the numerator. Therefore, one concludes that when $M_{S}$
and/or $M_{SS}$ go to zero, TRS diverges.

In order to elaborate the efficiency of the newly proposed metric
and shortcomings of the previously proposed metrics, we study two
cases of the BI black holes in three and four dimensions. We also
discuss linear electrodynamics cases and give a comment for
neutral solutions.

\section{Black hole solutions of Einstein gravity with a BI source}

The $d$-dimensional action of Einstein-BI gravity in the presence of
cosmological constant is given by \cite{Hendi}
\begin{equation}
I=-\frac{1}{16\pi }\int d^{d}x\sqrt{-g}\left[ R-2\Lambda +L(F)\right] ,
\label{Action}
\end{equation}
where $R$ is the Ricci scalar and $\Lambda $ refers to the
(negative) cosmological constant. Also, $L(F)$ is the Lagrangian
of BI field as follows \cite{BI}
\begin{equation}
L(F)=4\beta ^{2}\left( 1-\sqrt{1+\frac{F}{2\beta ^{2}}}\right) ,
\end{equation}
where $\beta $ is called the nonlinearity parameter, the Maxwell invariant $%
F=F_{\mu \nu }F^{\mu \nu }$ in which $F_{\mu \nu }=\partial _{\mu
}A_{\nu }-\partial _{\nu }A_{\mu }$ is the electromagnetic field
tensor and $A_{\mu } $ is the gauge potential. The static BI black
hole solutions can be obtained with the following $d$-dimensional
metric \cite{Hendi,Myung,Fernando,Dehghani}
\begin{equation}
ds^{2}=-f(r)dt^{2}+\frac{dr^{2}}{f(r)}+r^{2}d\Omega^{2},
\label{Metric}
\end{equation}%
where $d\Omega^{2}$ represents the line element of $r=constant$
and $t=constant$ hypersurfaces with volume $V_{d-2}$. For three
and four dimensional (3D and 4D) spacetimes, $V_{d-2}$ is equal to
$2\pi$ and $4\pi$, respectively, and $d\Omega^{2}$ can be written
with the following explicit forms
\begin{equation}
d\Omega ^{2}=\left\{
\begin{array}{cc}
d\theta ^{2}, & 3D \vspace{0.2cm} \\
d\theta ^{2}+\sin ^{2}\theta d\varphi ^{2}, & 4D%
\end{array}
\right. .
\end{equation}

The consistent metric functions $f(r)$ for two cases of three and four
dimensional black holes are given by \cite{Hendi,Myung,Fernando,Dehghani}
\begin{equation}
f(r)=\left\{
\begin{array}{cc}
-m-\Lambda r^{2}+2r^{2}\beta ^{2}\left( 1-\Gamma \right) +q^{2}\left[ 1-2\ln
\left( \frac{r\left( 1+\Gamma \right) }{2l}\right) \right]  & 3D\vspace{0.3cm%
} \\
1-\frac{m}{r}-\frac{\Lambda r^{2}}{3}+\frac{2\beta ^{2}}{3}r^{2}(1-\Gamma )+%
\frac{4q^{2}}{3r^{2}}\ _{2}\mathcal{F}_{1}\left( \left[
\frac{1}{2},\frac{1}{4}\right]
,\left[ \frac{5}{4}\right] ,1-\Gamma ^{2}\right)  & 4D%
\end{array}%
\right. ,  \label{metricfunc}
\end{equation}%
where $\Gamma =\sqrt{1+\frac{q^{2}}{r^{2(d-2)}\beta ^{2}}}$, $\
_{2}\mathcal{F}_{1}$ is hypergeometric function, $m$ and $q$ are
integration constants which are related to mass parameter and the
electric charge of the black holes, respectively. The entropy and
the electric charge of the mentioned BI black hole solutions were
obtained before \cite{Hendi,Myung,Fernando,Dehghani}
\begin{equation}
S=\frac{V_{d-2}\;r_{+}^{d-2}}{4},  \label{Entropy}
\end{equation}%
\begin{equation}
Q=\frac{V_{d-2}\;q}{4\pi },  \label{Q}
\end{equation}%
where $r_{+}$ denotes the outer (event) horizon of black holes
which is the largest real positive root of metric function,
$f(r)|_{r=r_{+}}=0$.

Now, we write the quasi-local mass (per unit volume $V_{d-2}$) as
a function of extensive parameters ($S$ and $Q$) to discuss phase
transition. One finds \cite{Hendi,Myung,Fernando,Dehghani}

\begin{equation}
M=\frac{(d-2)}{16\pi }\times \left\{
\begin{array}{cc}
\frac{4\left( \pi ^{2}Q^{2}-\Lambda S^{2}\right) +8S^{2}\beta ^{2}\left( 1-%
\mathcal{H}\right) -8\pi ^{2}Q^{2}\ln \left( \frac{S}{\pi l}\left( 1+%
\mathcal{H}\right) \right) }{\pi ^{2}} & 3D\vspace{0.3cm} \\
\frac{3S\left( \pi -\Lambda S\right) +2S^{2}\beta ^{2}\left( 1-\mathcal{H}%
\right) +4\pi ^{2}Q^{2}\mathfrak{F}}{3\pi ^{2}\sqrt{\frac{S}{\pi }}} & 4D%
\end{array}%
\right. ,  \label{mass}
\end{equation}%
where $\mathfrak{F}=\mathcal{F}\left( \left[ \frac{1}{2},\frac{1}{4}\right] ,%
\left[ \frac{5}{4}\right] ,-\frac{\pi ^{2}Q^{2}}{S^{2}\beta ^{2}}\right) $
and $\mathcal{H}=\sqrt{1+\frac{\pi ^{2}Q^{2}}{S^{2}\beta ^{2}}}$. Regarding
Eq. (\ref{heat}), one finds that the heat capacity for three and four
dimensional BI black holes can be written as
\begin{equation}
C_{Q}=\left\{
\begin{array}{cc}
\frac{S^{2}\left( 1+\mathcal{H}\right) \left[ \Lambda \beta ^{2}S^{2}\left(
1+\mathcal{H}\right) +\pi ^{2}Q^{2}\left( \Lambda +2\beta ^{2}\mathcal{H}%
\right) \right] }{2\beta ^{2}S\left( \Lambda S^{2}-\pi ^{2}Q^{2}\right)
\left( 1+\mathcal{H}\right) +\pi ^{2}Q^{2}\Lambda \left( 2+\mathcal{H}%
\right) } & 3D\vspace{0.5cm} \\
-\frac{2\beta ^{2}\mathcal{H}^{2}S^{3}\left\{ \left[ \Sigma _{+}+4Q^{2}\pi
^{2}(\mathfrak{F}+4Q^{2}\mathfrak{F}^{\prime })\right] \mathcal{H}+2Q^{2}\pi
^{2}+6\beta ^{2}S^{2}\right\} }{\beta ^{2}\mathcal{H}^{3}S^{2}\left[ \Sigma
_{-}+4Q^{2}\pi ^{2}(3\mathfrak{F}+32Q^{2}\mathfrak{F}^{\prime }+16Q^{4}%
\mathfrak{F}^{\prime \prime })\right] \mathcal{H}+2Q^{4}\pi ^{4}-6\beta
^{2}S^{2}(\beta ^{2}S^{2}+2Q^{2}\pi ^{2})} & 4D%
\end{array}%
\right. ,  \label{HCap}
\end{equation}%
where $\Sigma _{\pm }=-3S\pi \pm $ $3S^{2}(\Lambda -2\beta ^{2})$. It is
notable that the prime and double prime are the first and second derivatives
with respect to $Q^{2}$, respectively.

\begin{figure}[tbp]
$%
\begin{array}{cc}
\epsfxsize=6cm \epsffile{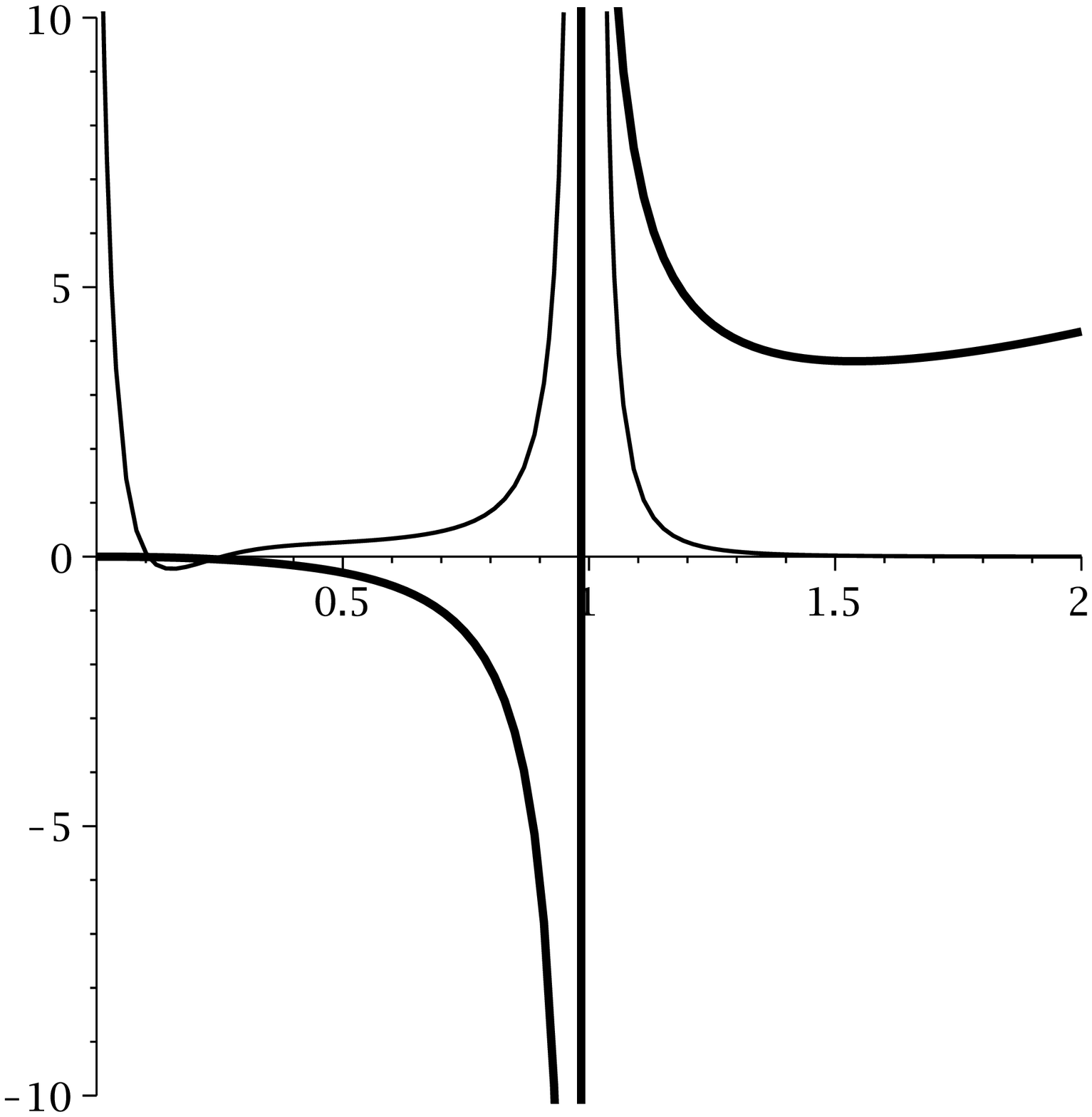} & \epsfxsize=6cm %
\epsffile{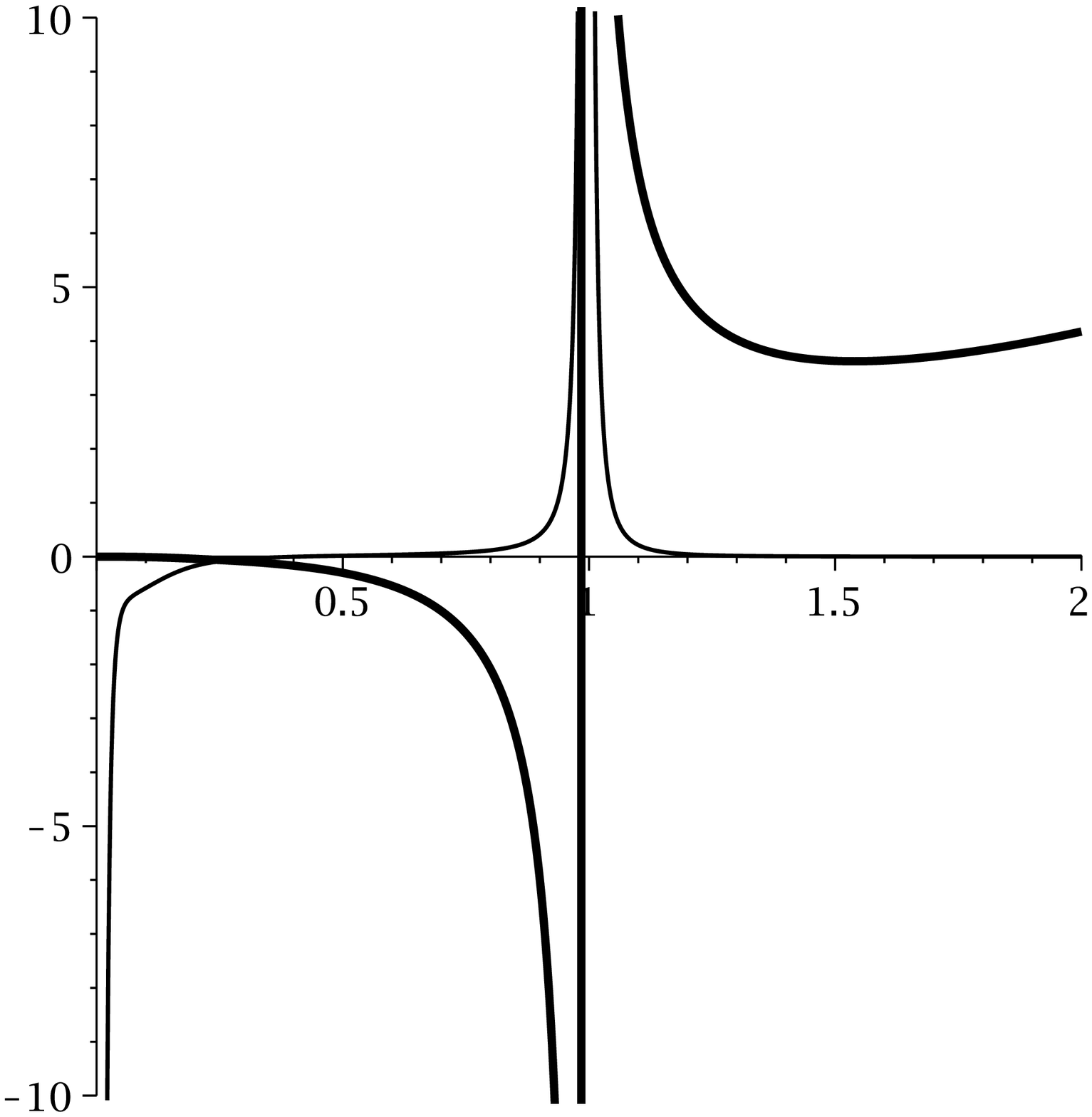} \\
\epsfxsize=6cm \epsffile{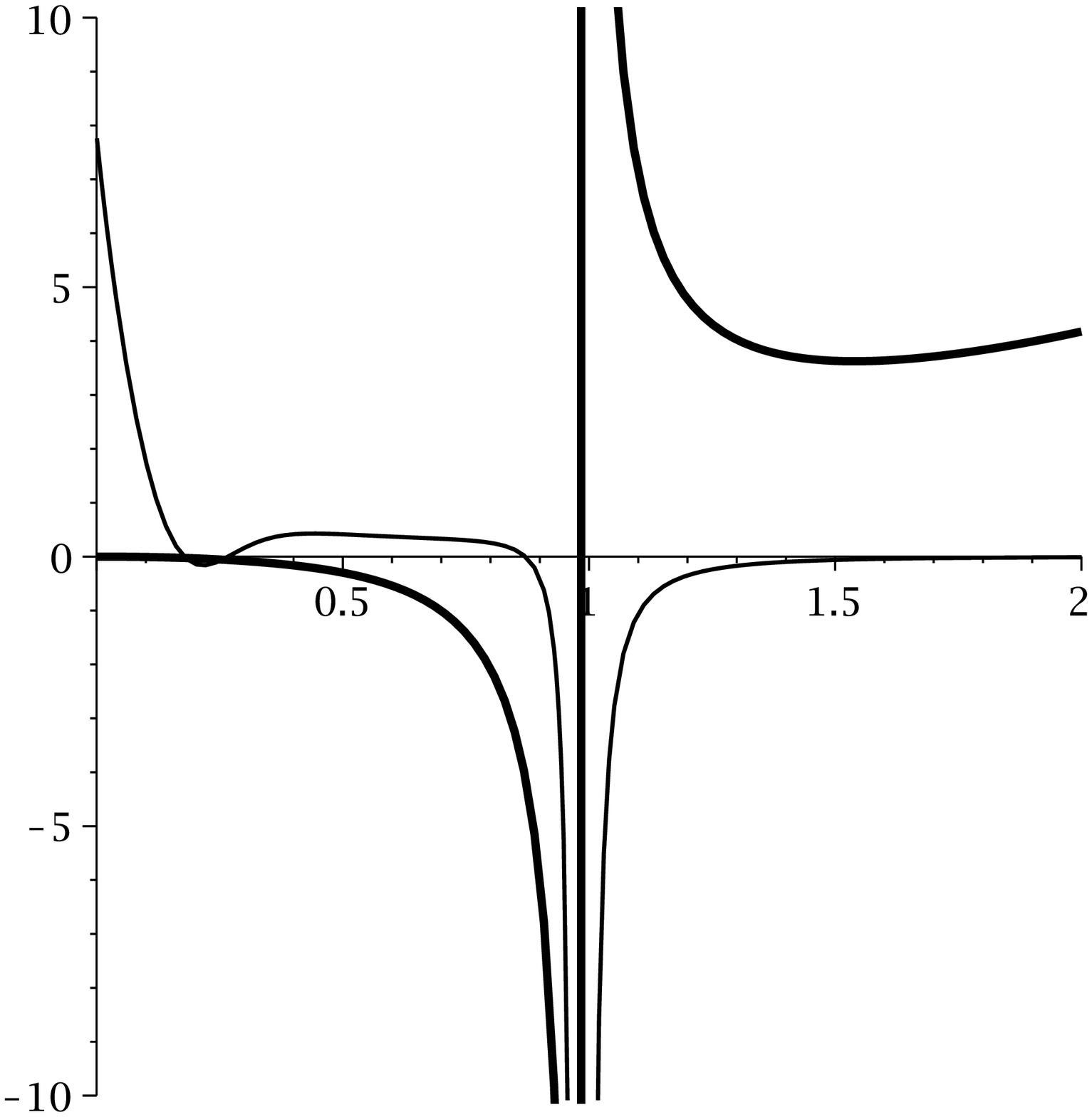} & \epsfxsize=6cm %
\epsffile{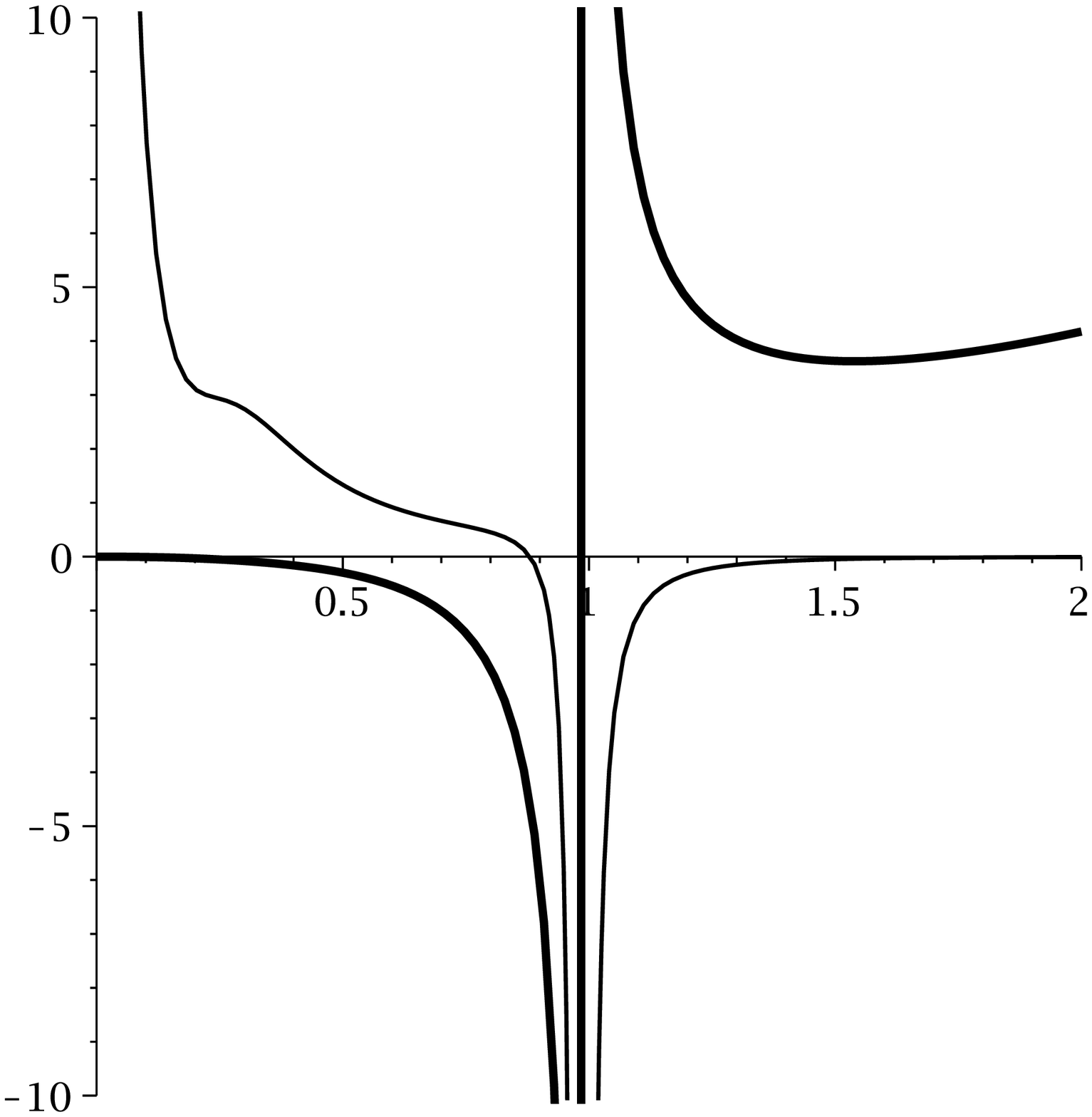}%
\end{array}
$%
\caption{\textbf{BI case:} $\mathcal{R}$ and $C_{Q}$ versus $r_{+}$ for $l=1$%
, $\Lambda =-1$, $\protect\beta =1$, $d=4$ and $q=0.1$. \newline
TRS (continuous line) and heat capacity (bold line) for the
Weinhold metric (left-up panel), the Ruppeiner metric (right-up
panel), the Quevedo metric for case I (left-down panel) and the
Quevedo metric for case II (right-down panel). } \label{Fig1}
\end{figure}

\begin{figure}[tbp]
$%
\begin{array}{cc}
\epsfxsize=6cm \epsffile{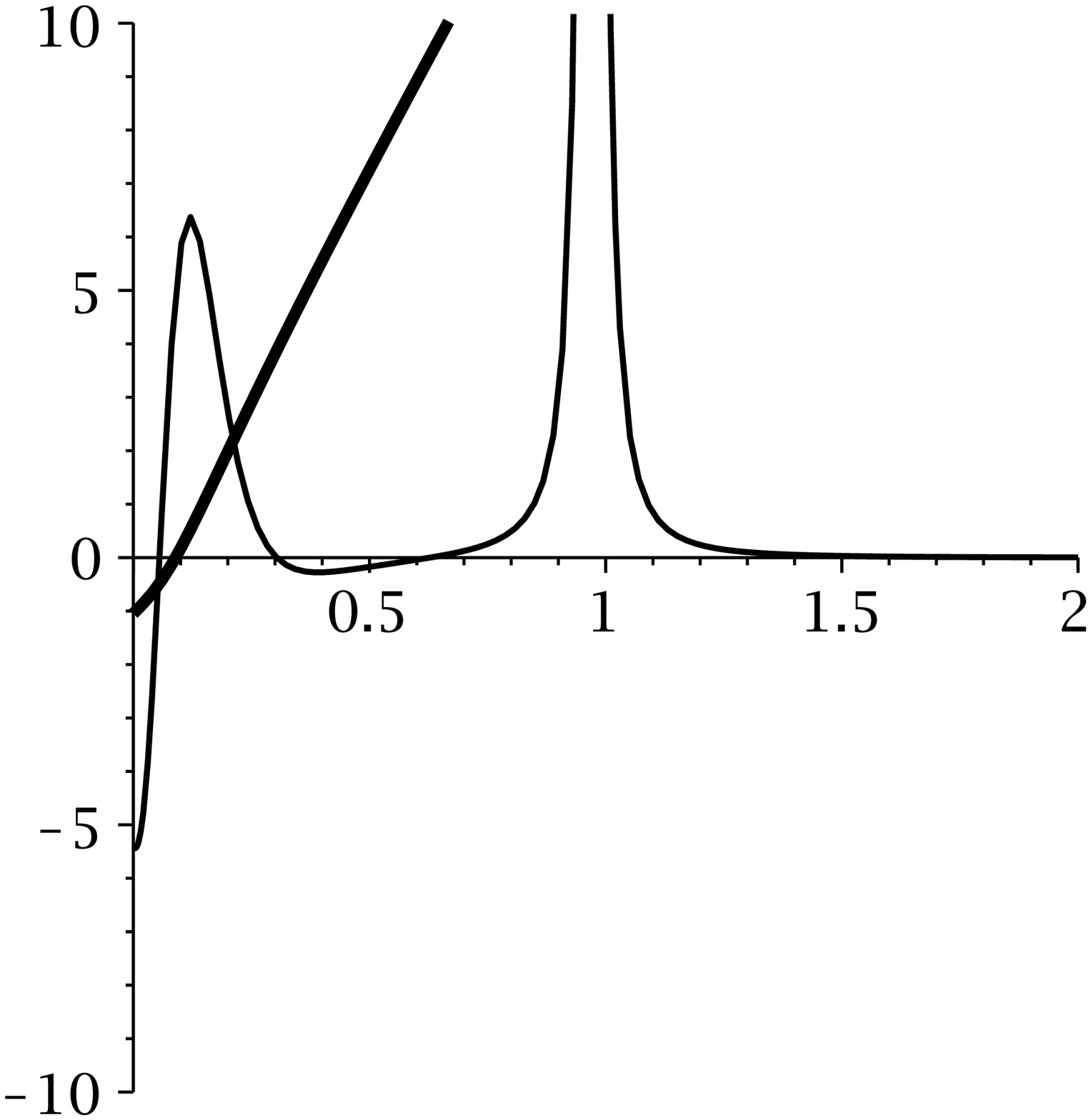} & \epsfxsize=6cm %
\epsffile{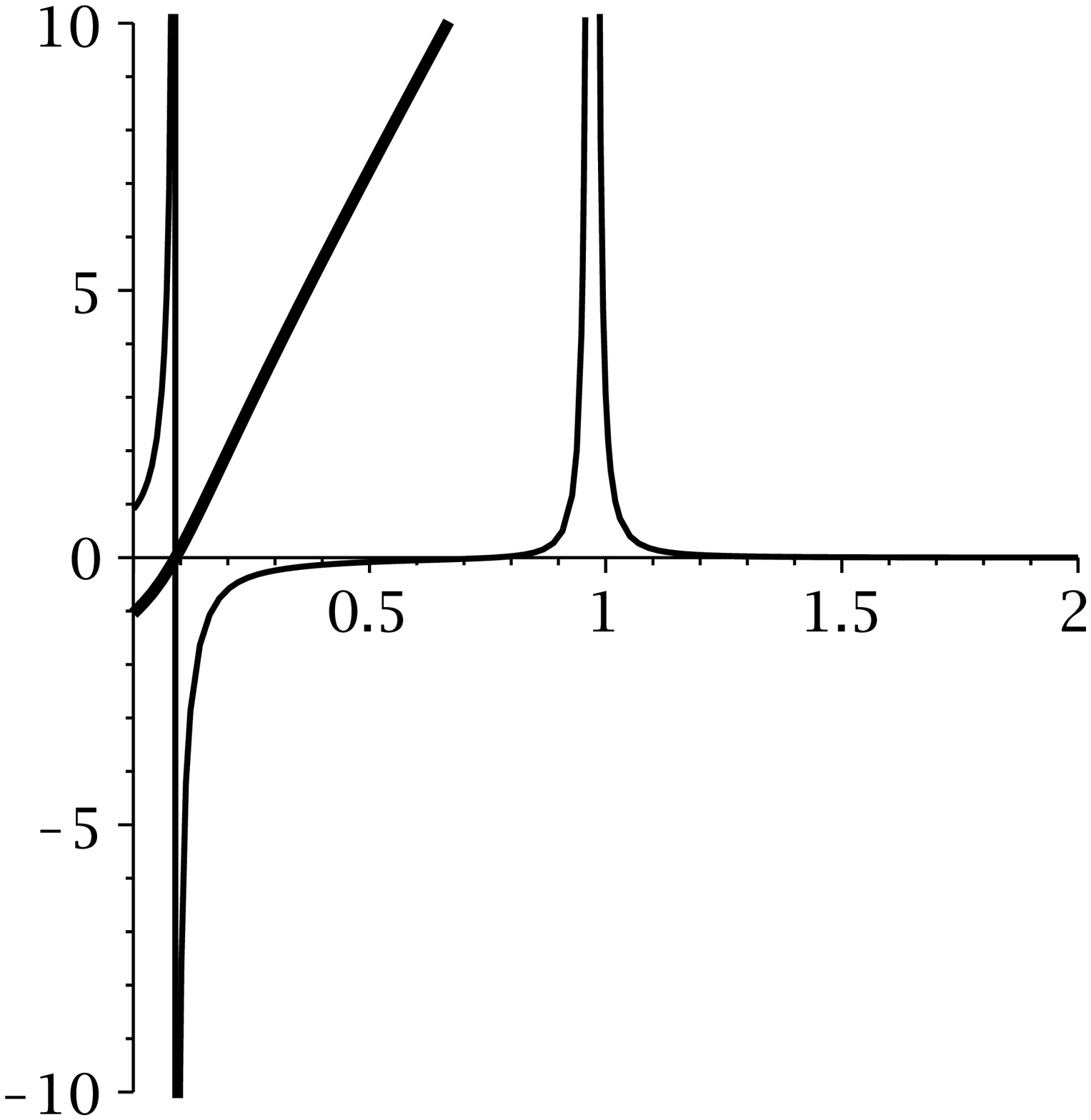} \\
\epsfxsize=6cm \epsffile{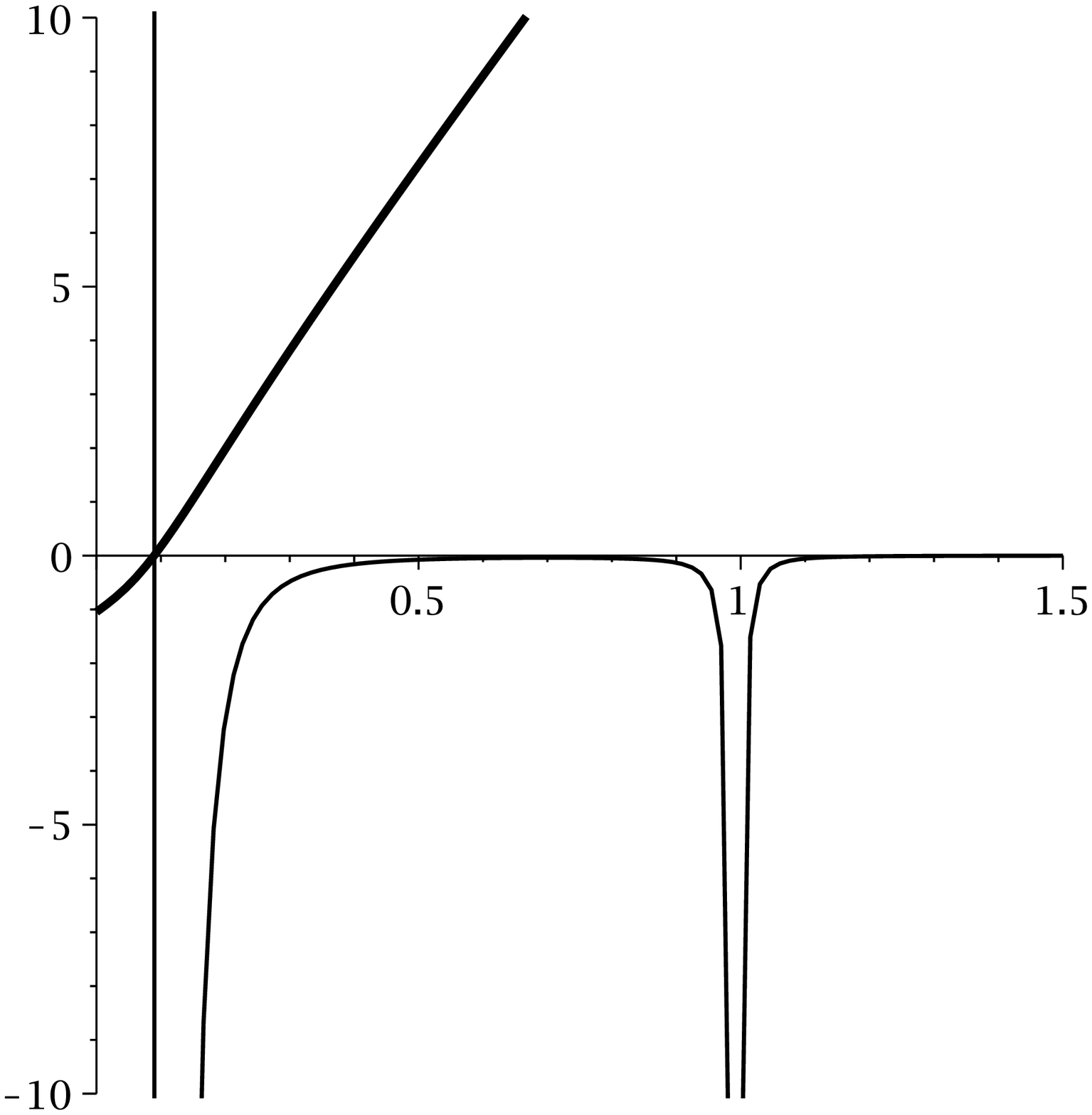} & \epsfxsize=6cm %
\epsffile{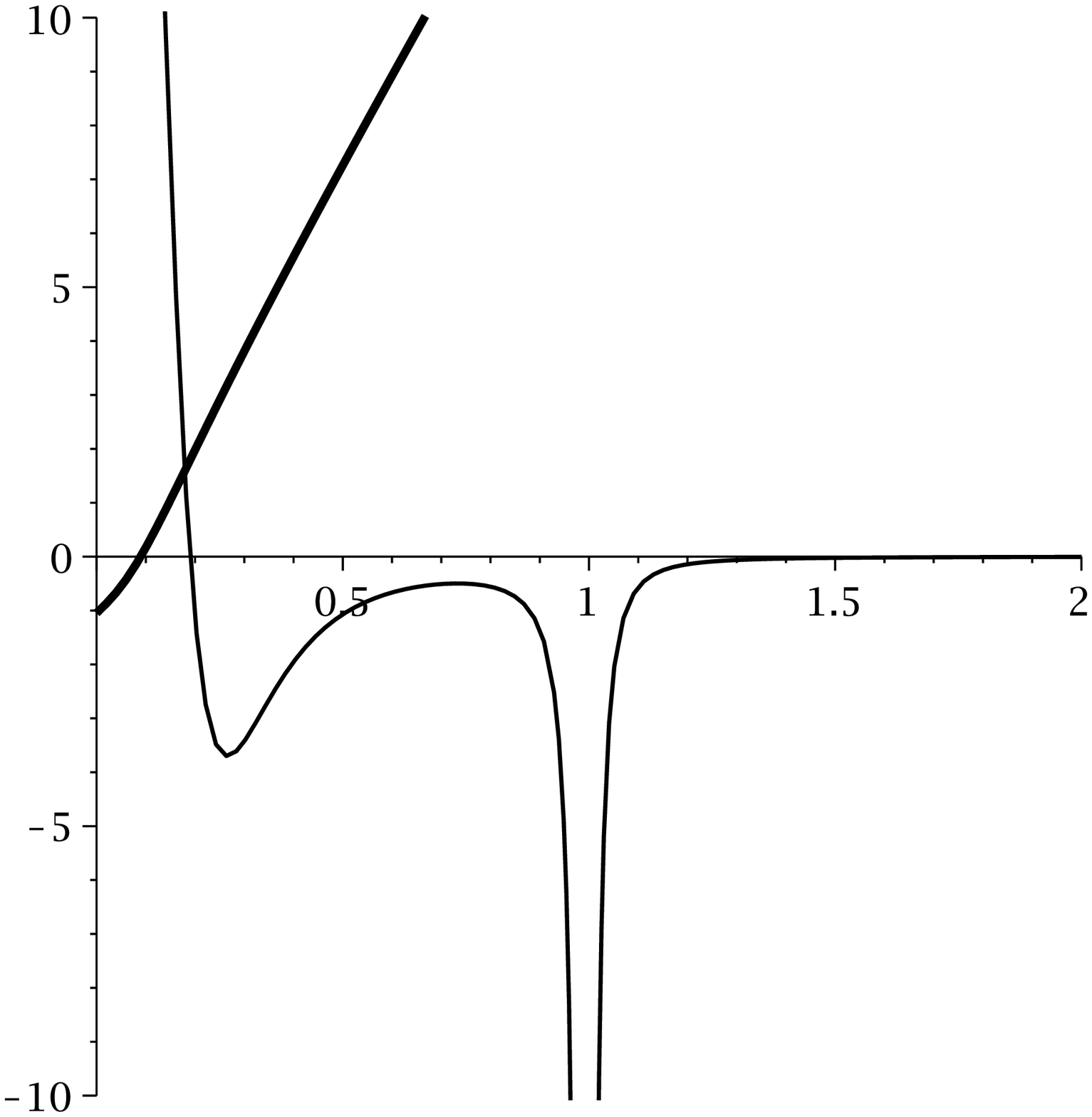}%
\end{array}
$%
\caption{\textbf{BI case:} $\mathcal{R}$ and $C_{Q}$ versus $r_{+}$ for $l=1$%
, $\Lambda =-1$, $\protect\beta =1$, $d=3$ and $q=0.1$. \newline
TRS (continuous line) and heat capacity (bold line) for the
Weinhold metric (left-up panel), the Ruppeiner metric (right-up
panel), the Quevedo metric for case I (left-down panel) and the
Quevedo metric for case II (right-down panel). } \label{Fig2}
\end{figure}

\begin{figure}[tbp]
$%
\begin{array}{cc}
\epsfxsize=6cm \epsffile{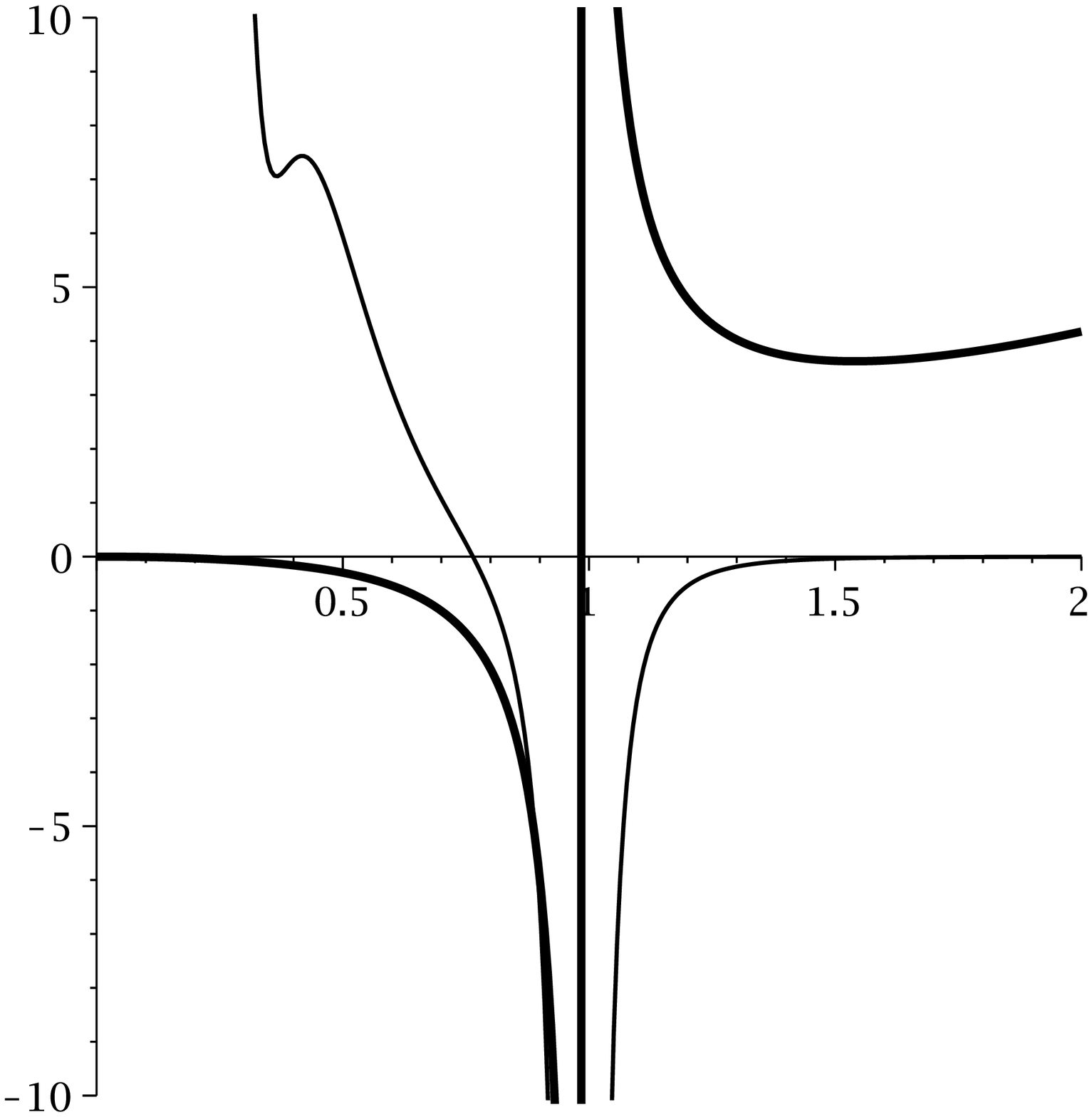} & \epsfxsize=6cm %
\epsffile{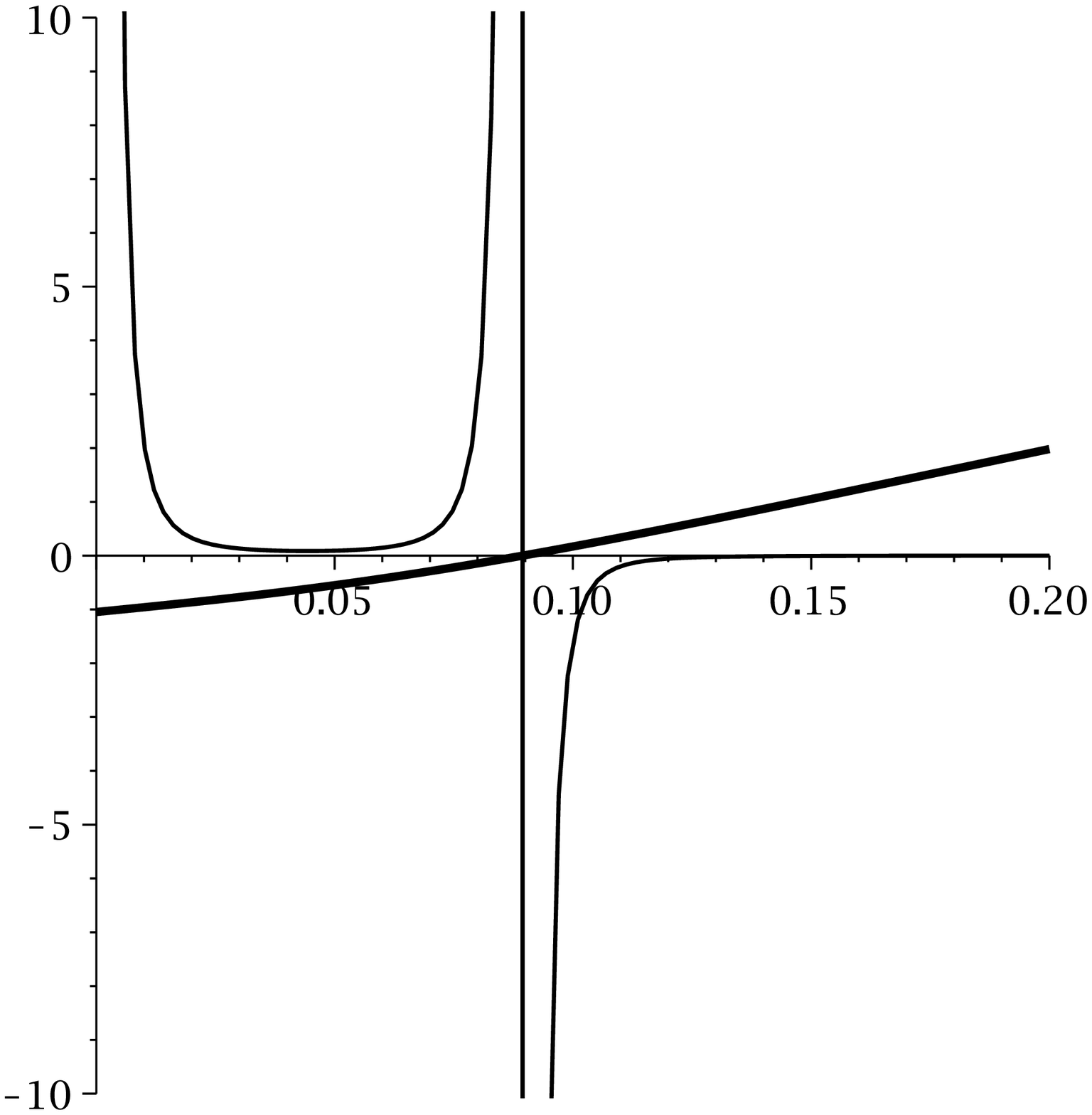}%
\end{array}
$%
\caption{\textbf{BI case:} $\mathcal{R}$ and $C_{Q}$ versus $r_{+}$ for $l=1$%
, $\Lambda =-1$, $\protect\beta =1$, $q=0.1$.\newline TRS
(continuous line) and heat capacity (bold line) for the new metric
with $d=4$ (left panel) and $d=3$ (right panel)} \label{Fig3}
\end{figure}

Here, we regard Eq. (\ref{HCap}) and also all obtained TRS in
various mentioned methods ( Eqs. (\ref{Wein}), (\ref{Rupp}),
(\ref{Quev}) and (\ref{newmetric}) ) to investigate their
coincidences. To do so, we plot some figures for three and four
dimensional BI black hole solutions (see Figs.
\ref{Fig1}-\ref{Fig3}). We find that for $4$-dimensional BI
solutions the Ruppeiner, the Weinhold, both methods of the Quevedo
and also our new introduced metric have divergencies where the
heat capacity diverges (see Fig. \ref{Fig1} and left panel of Fig.
\ref{Fig3} for more details). In other words, applying the
mentioned methods for these solutions leads to coincidence between
divergencies of TRS and phase transitions of the type two, and
therefore, we conclude that for these solutions ($4$-dimensional
BI black holes) extra terms in Quevedo metrics, have no
contribution to the divergencies of TRS.

Next, we take into account three dimensional solutions. We find
that for all methods of the Weinhold, the Ruppeiner and both cases
of the Quevedo metrics (Eqs. (\ref{Wein}), (\ref{Rupp}) and
(\ref{Quev})), there are extra divergencies in plotted figures of
TRS which come from the contributions of extra term (see Fig.
\ref{Fig2}). While for the TRS of our new metric (Eq.
(\ref{newmetric})) all divergencies coincide with the phase
transitions of heat capacity (see Fig. \ref{Fig3} (right panel)
for more details).

Another important property of new metric is the behavior of TRS very close
to the phase transition points. As Fig. \ref{Fig3} confirms, the sign of the
TRS before and after the divergence point for these two types of the phase
transitions is different. If the phase transition is related to the
vanishing heat capacity (type one), we see a change of sign for TRS before
and after of the corresponding singular point. While for the divergence
point of the heat capacity (type two), TRS has the same sign for the left
and right sides of this point. Therefore, this characteristic behavior
enables one to distinguish these two types of phase transitions from each
other.

\subsection{Linear Case: Maxwell solutions}

In order to study the effect of electric charge and remove the influence of
nonlinearity parameter, we investigate the Maxwell solutions. We show that
these linear solutions elaborate the efficiency of our new metric. In order
to obtain Maxwell solutions, one can use series expansion of BI solutions
for large values of nonlinearity parameter $\beta $. Regarding the mentioned
results of BI solutions with straightforward series expansion, one can
obtain \cite{threeMaxwell,fourMaxwell}
\begin{equation}
f(r)=\left\{
\begin{array}{cc}
-m-\Lambda r^{2}-2q^{2}\ln \left( \frac{r}{l}\right)  & 3D\vspace{0.3cm} \\
1-\frac{m}{r}-\frac{\Lambda r^{2}}{3}+\frac{q^{2}}{r^{2}} & 4D%
\end{array}%
\right. .
\end{equation}

Since the entropy and electric charge of these black holes do not
depend on the nonlinearity parameter, one can obtain the same
results. Therefore, the finite mass (per unit volume $V_{d-2}$) of
the Einstein-Maxwell solutions can be written as a function of $S$
and $Q$ with the following explicit form
\begin{equation}
M=\frac{(d-2)}{16\pi }\times \left\{
\begin{array}{cc}
\frac{2\pi ^{2}Q^{2}\ln \left( \frac{\pi l}{2S}\right) -4\Lambda S^{2}}{\pi
^{2}} & 3D\vspace{0.3cm} \\
\frac{3\pi S+3\pi ^{2}Q^{2}-\Lambda S^{2}}{3\pi ^{2}\sqrt{\frac{S}{\pi }}} &
4D%
\end{array}%
\right. .  \label{massMax}
\end{equation}

Using Eqs. (\ref{heat}) and (\ref{massMax}), one can find the heat capacity
may be calculated as
\begin{equation}
C_{Q}\mathbf{=}\left\{
\begin{array}{cc}
-\frac{\left( \pi ^{2}Q^{2}+\Lambda S^{2}\right) S}{\pi ^{2}Q^{2}-\Lambda
S^{2}} & 3D\vspace{0.3cm} \\
-\frac{2S\left( \pi S-\Lambda S^{2}-\pi ^{2}Q^{2}\right) }{\Lambda S^{2}+\pi
S-3\pi ^{2}Q^{2}} & 4D%
\end{array}%
\right. .  \label{HC}
\end{equation}

\begin{figure}[tbp]
$%
\begin{array}{cc}
\epsfxsize=6cm \epsffile{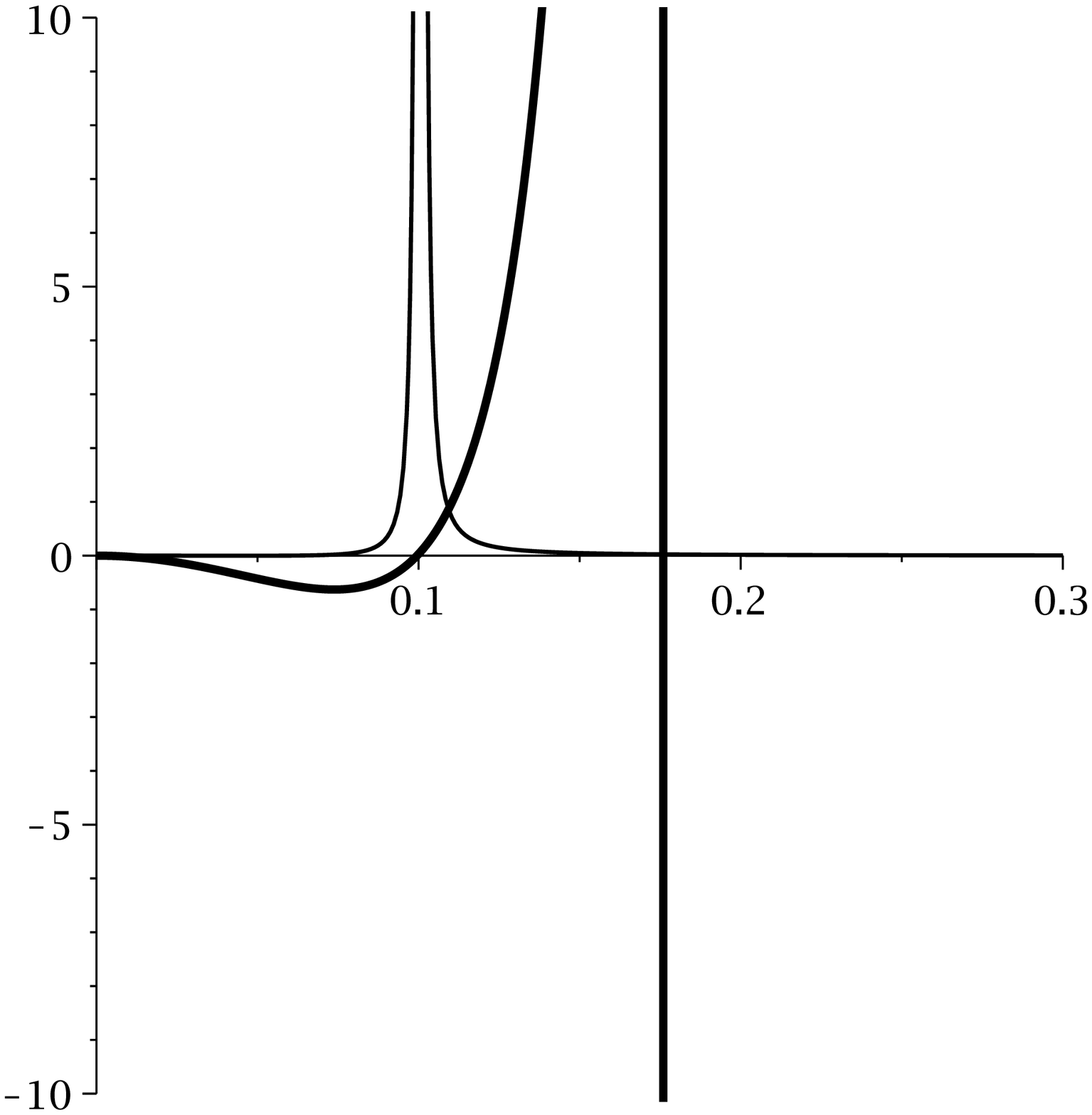} & \epsfxsize=6cm %
\epsffile{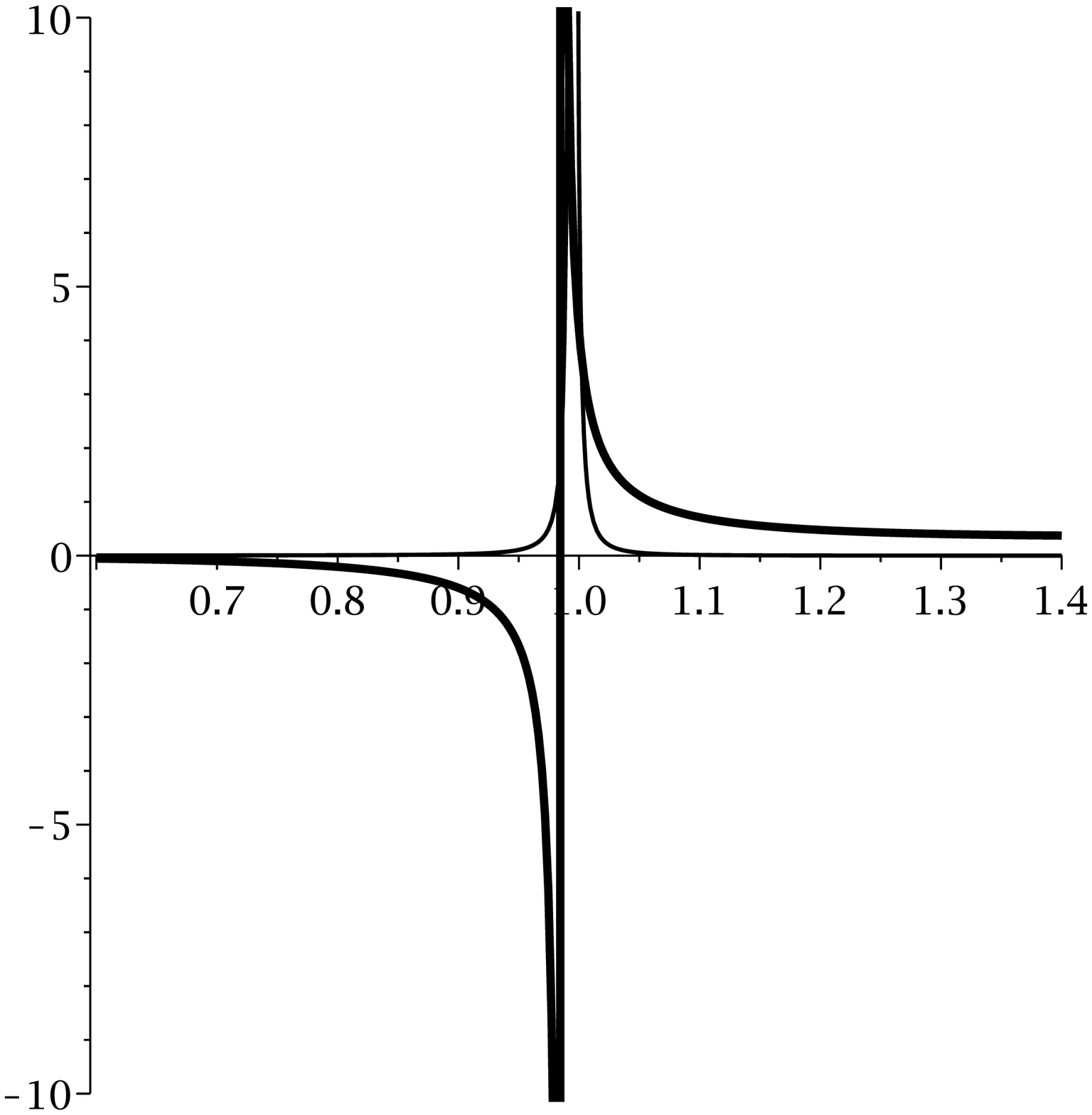} \\
\epsfxsize=6cm \epsffile{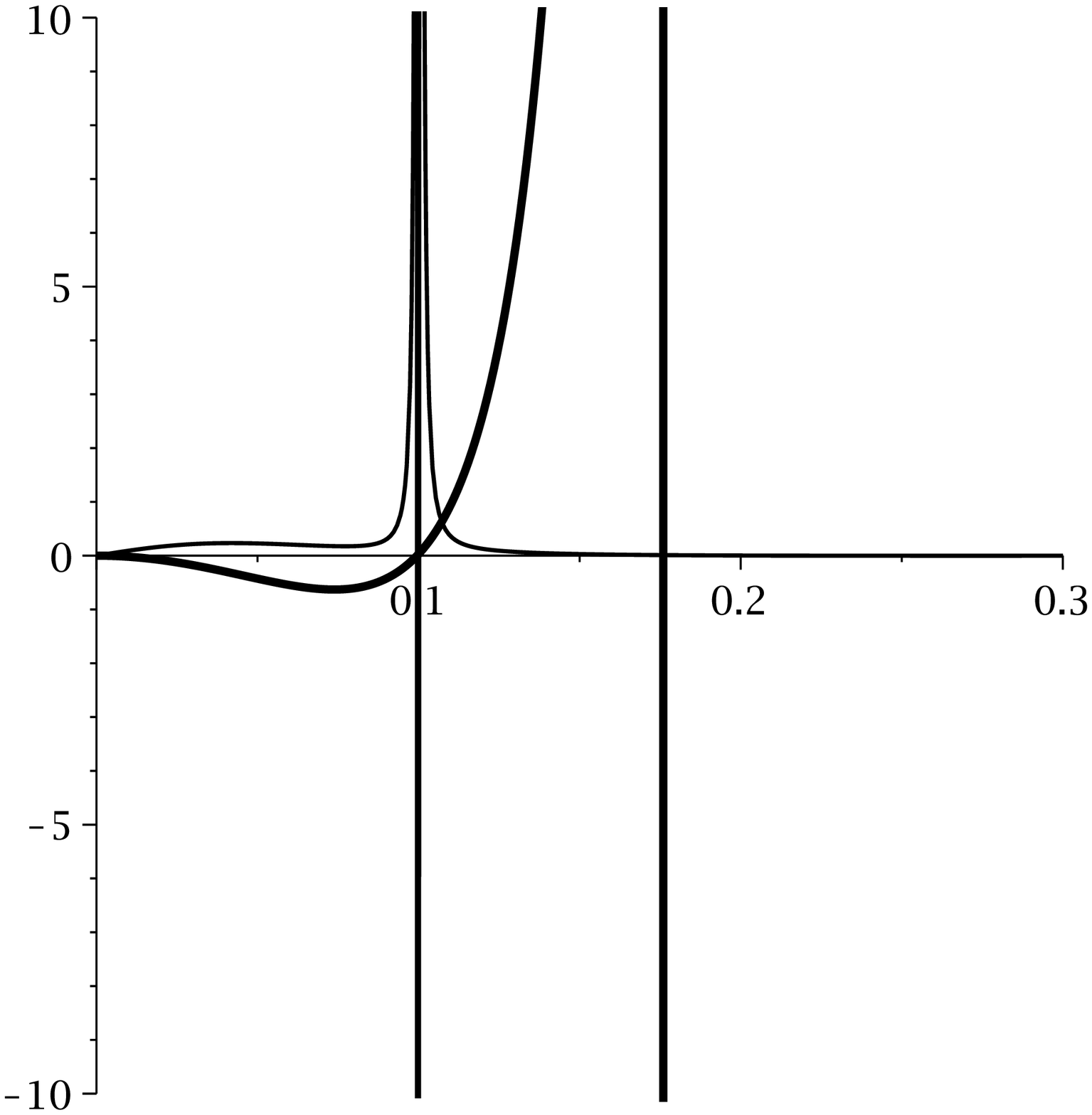} & \epsfxsize=6cm %
\epsffile{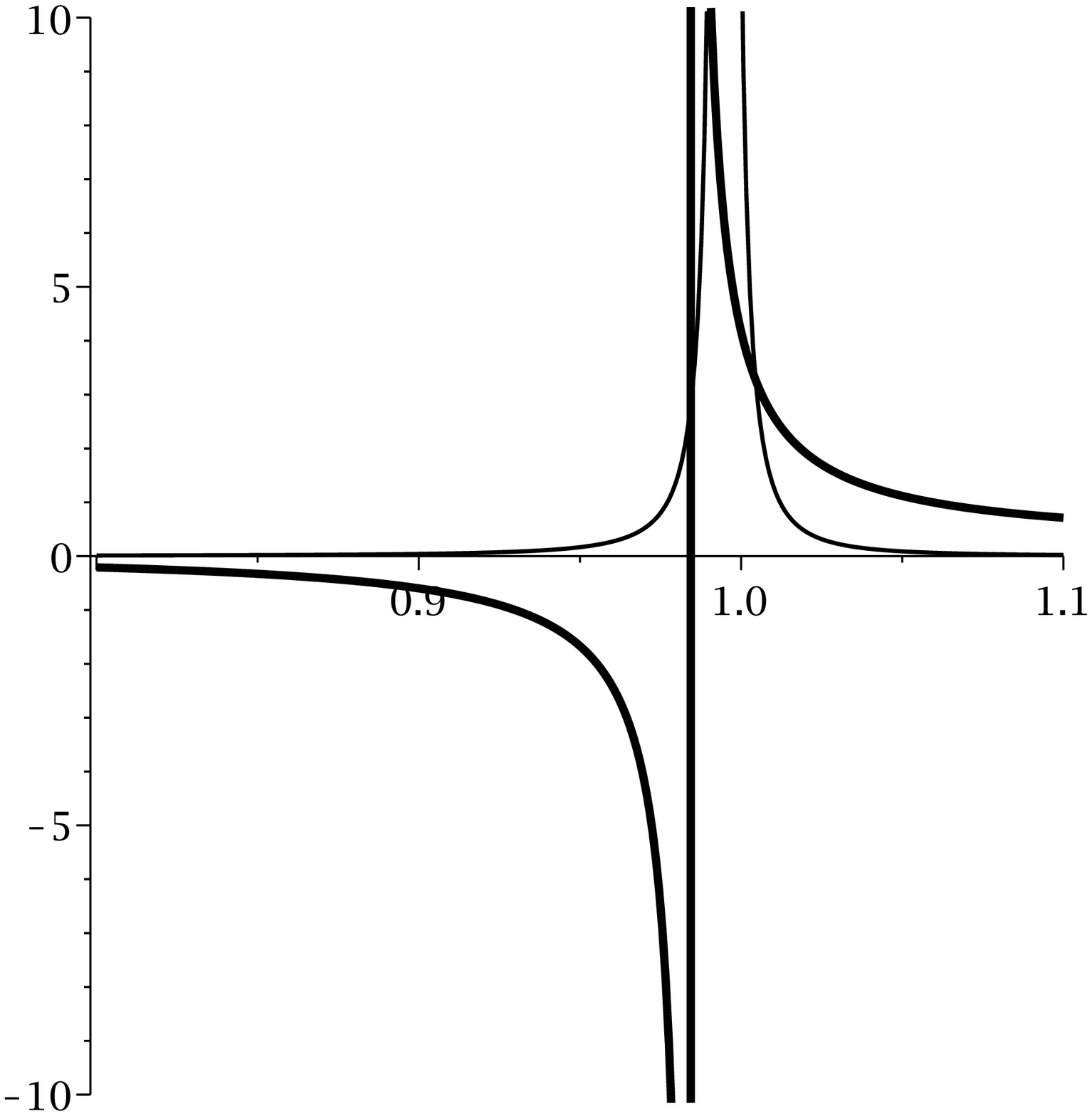}%
\end{array}
$%
\caption{\textbf{Maxwell case:} $\mathcal{R}$ and $C_{Q}$ versus
$r_{+}$ for $l=1$, $\Lambda =-1$, $d=4$ and $q=0.1$. \newline TRS
(continuous line) and heat capacity (bold line) for the Weinhold
metric (up panels) and the Ruppeiner metric (down panels). }
\label{Fig4}
\end{figure}

\begin{figure}[tbp]
$%
\begin{array}{cc}
\epsfxsize=6cm \epsffile{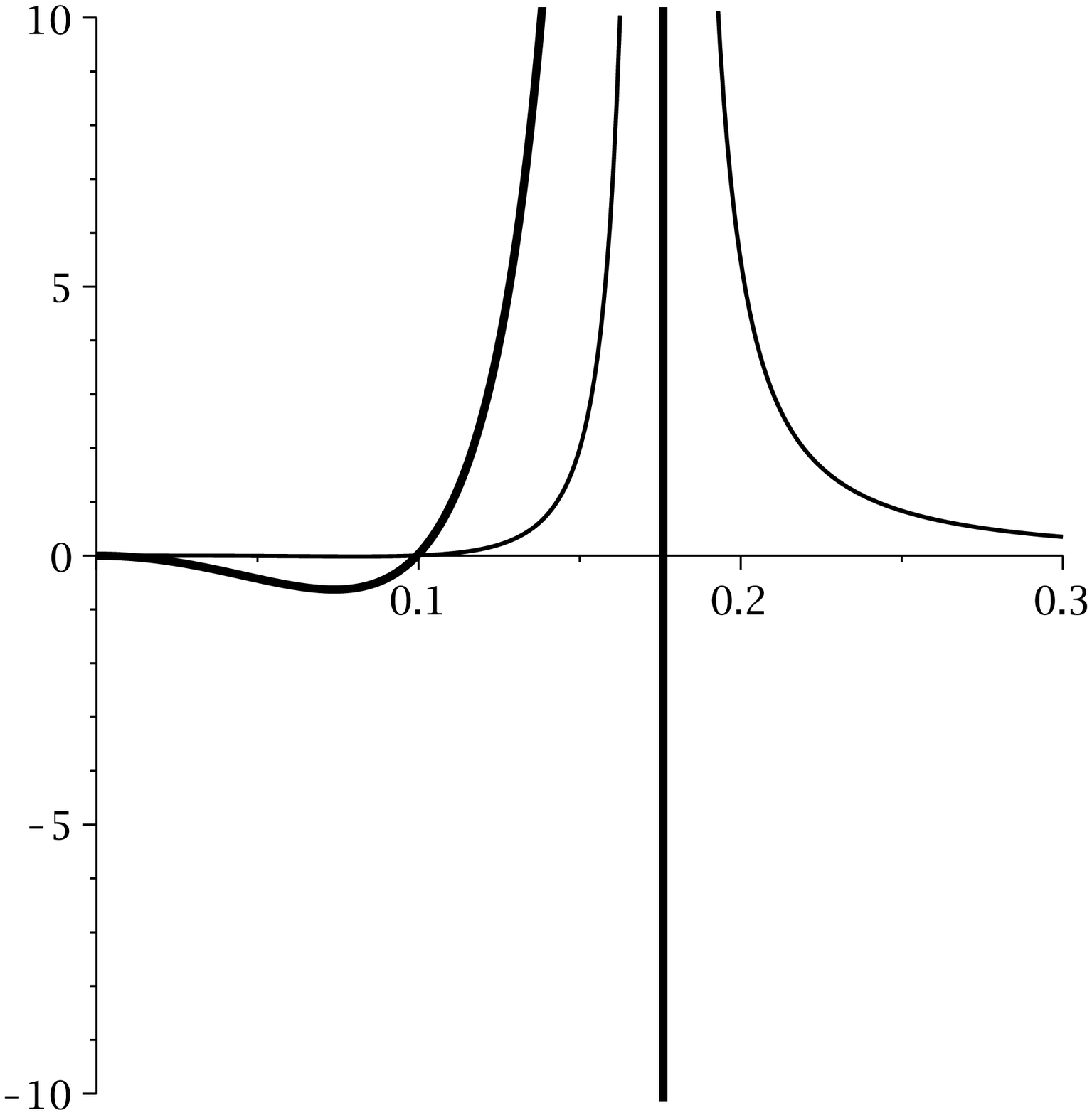} & \epsfxsize=6cm %
\epsffile{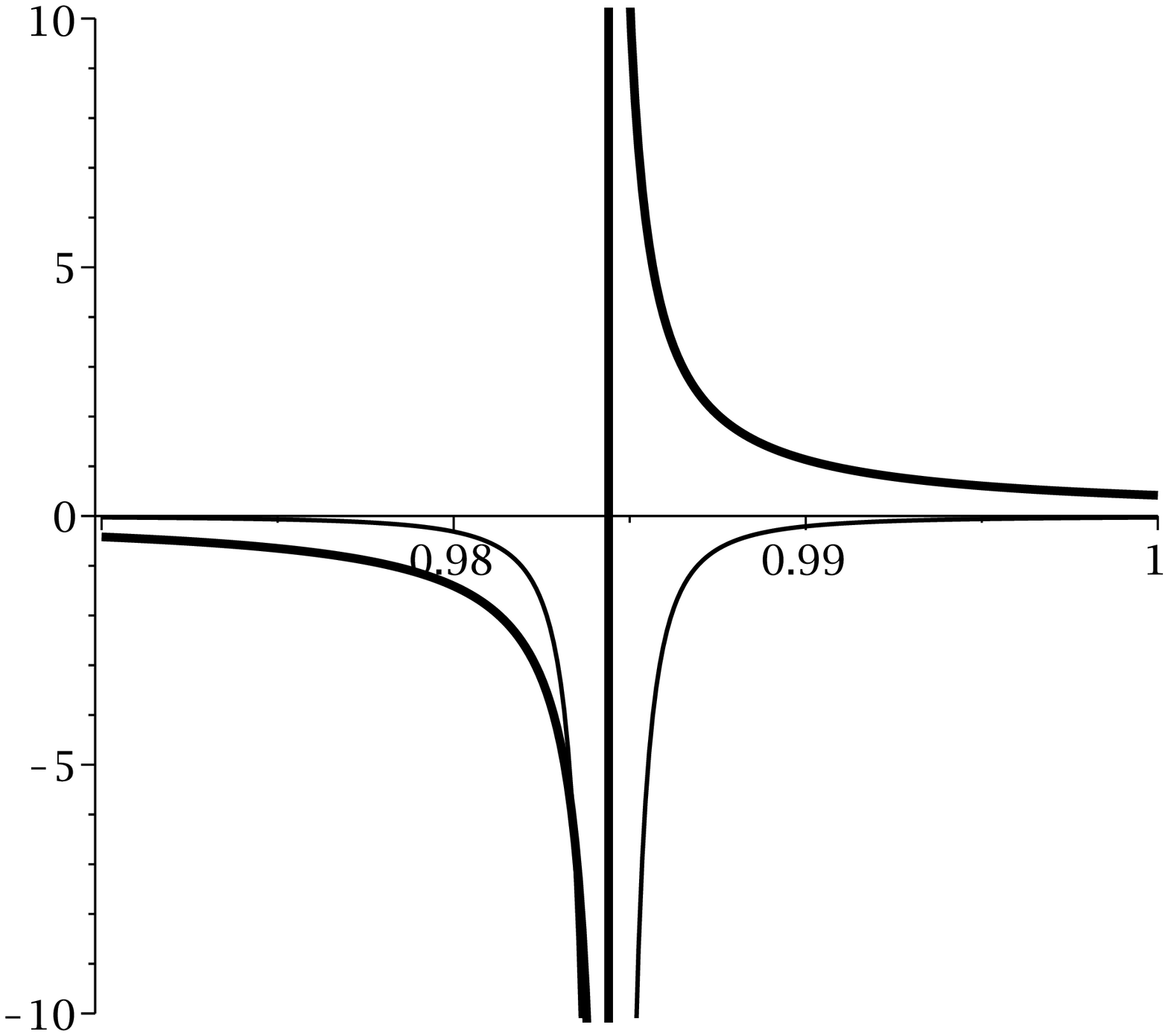} \\
\epsfxsize=6cm \epsffile{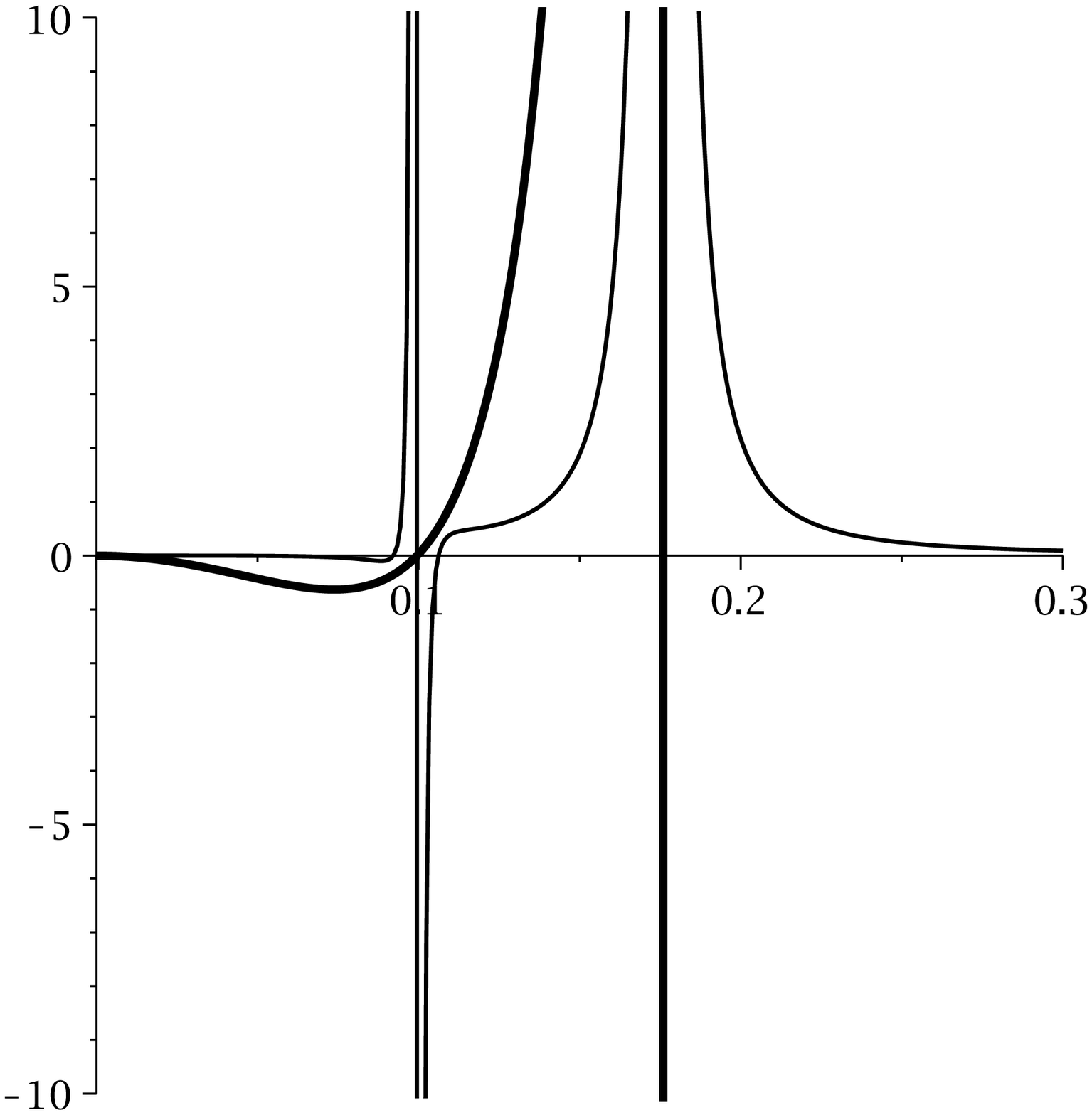} & \epsfxsize=6cm %
\epsffile{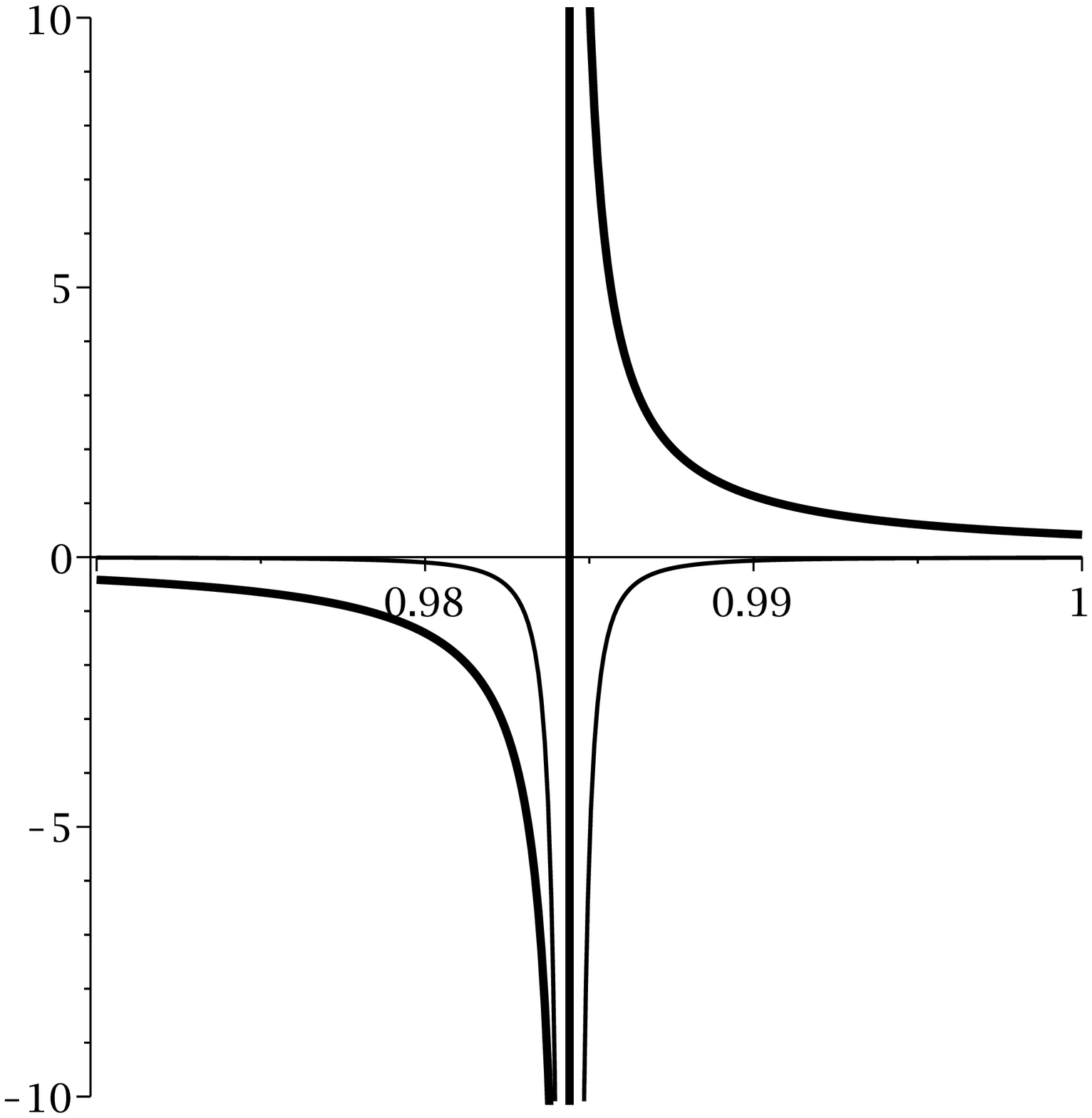}%
\end{array}
$%
\caption{\textbf{Maxwell case:} $\mathcal{R}$ and $C_{Q}$ versus
$r_{+}$ for $l=1$, $\Lambda =-1$, $d=4$ and $q=0.1$. \newline TRS
(continuous line) and heat capacity (bold line) for the Quevedo
metric for case I (up panels) and the Quevedo metric for case II
(down panels). } \label{Fig5}
\end{figure}

\begin{figure}[tbp]
$%
\begin{array}{cc}
\epsfxsize=6cm \epsffile{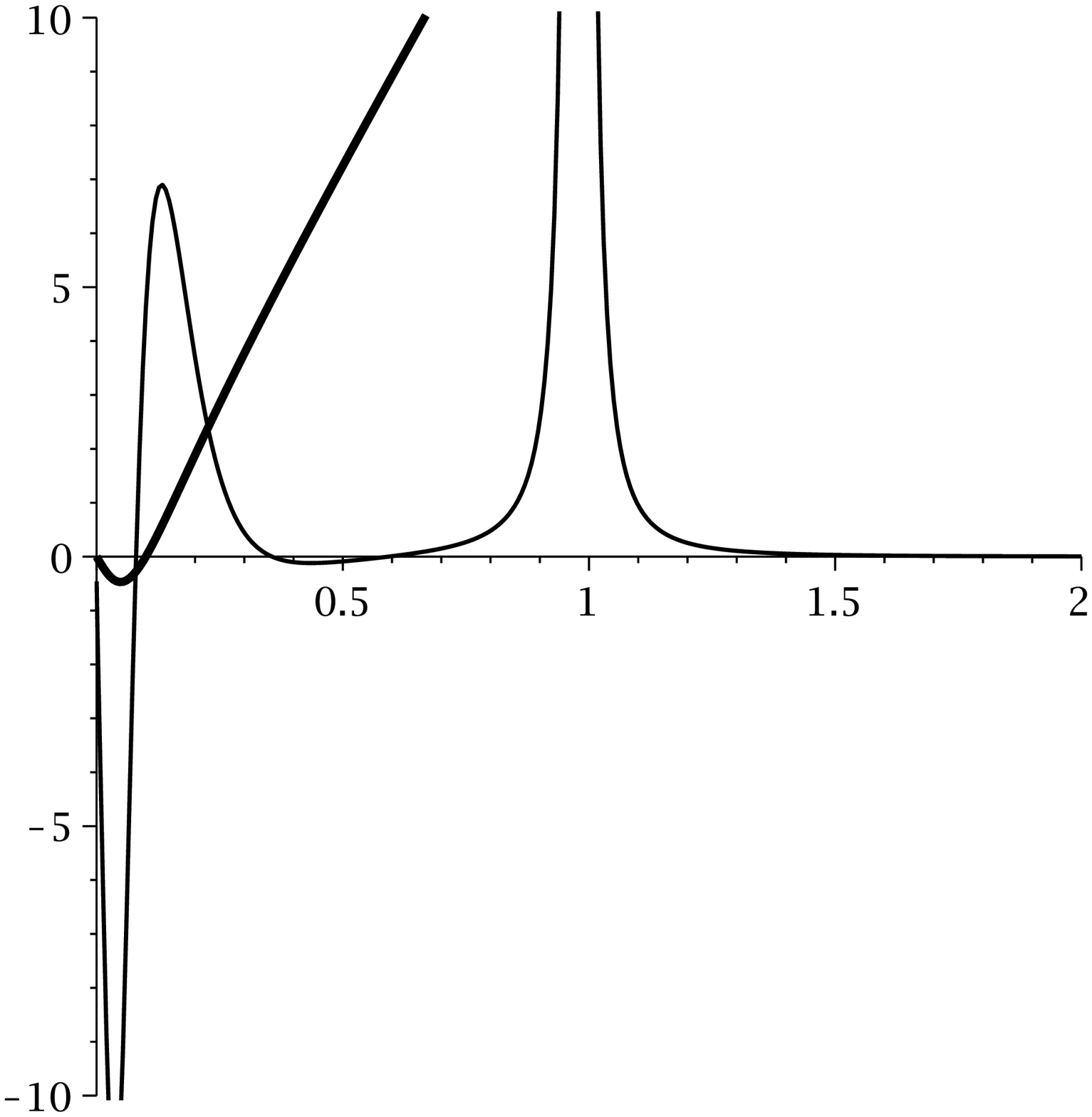} & \epsfxsize=6cm %
\epsffile{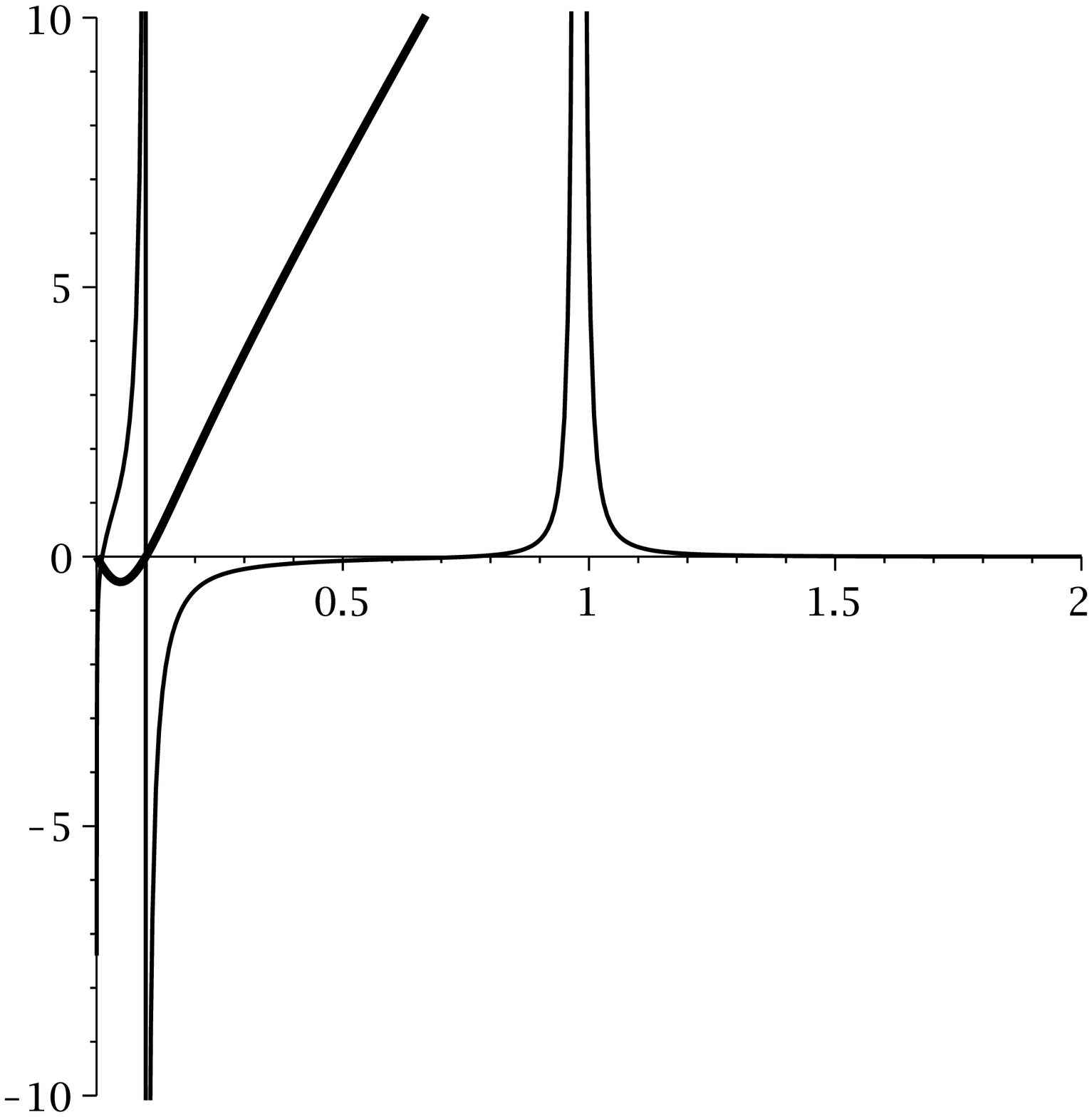} \\
\epsfxsize=6cm \epsffile{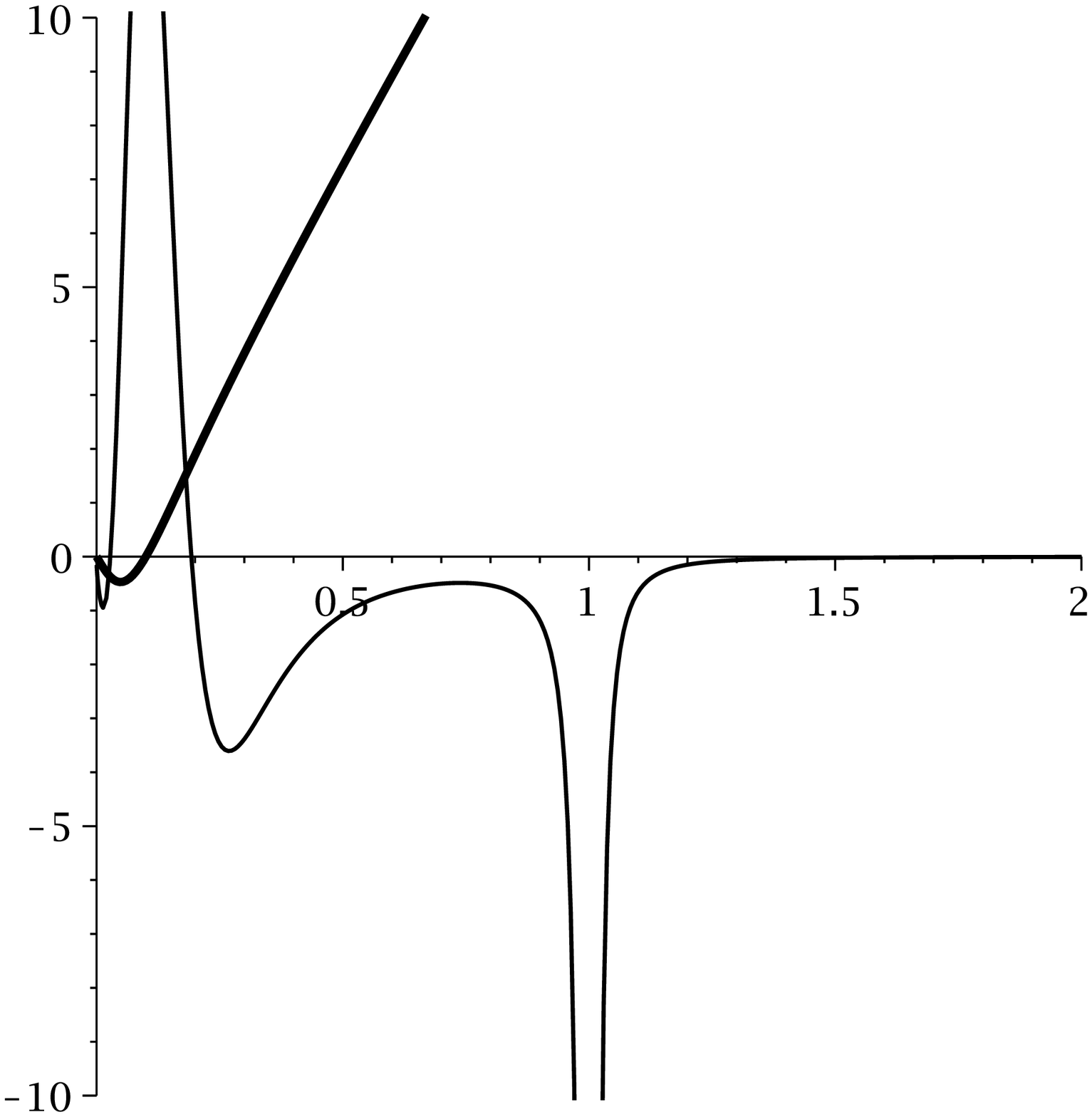} & \epsfxsize=6cm %
\epsffile{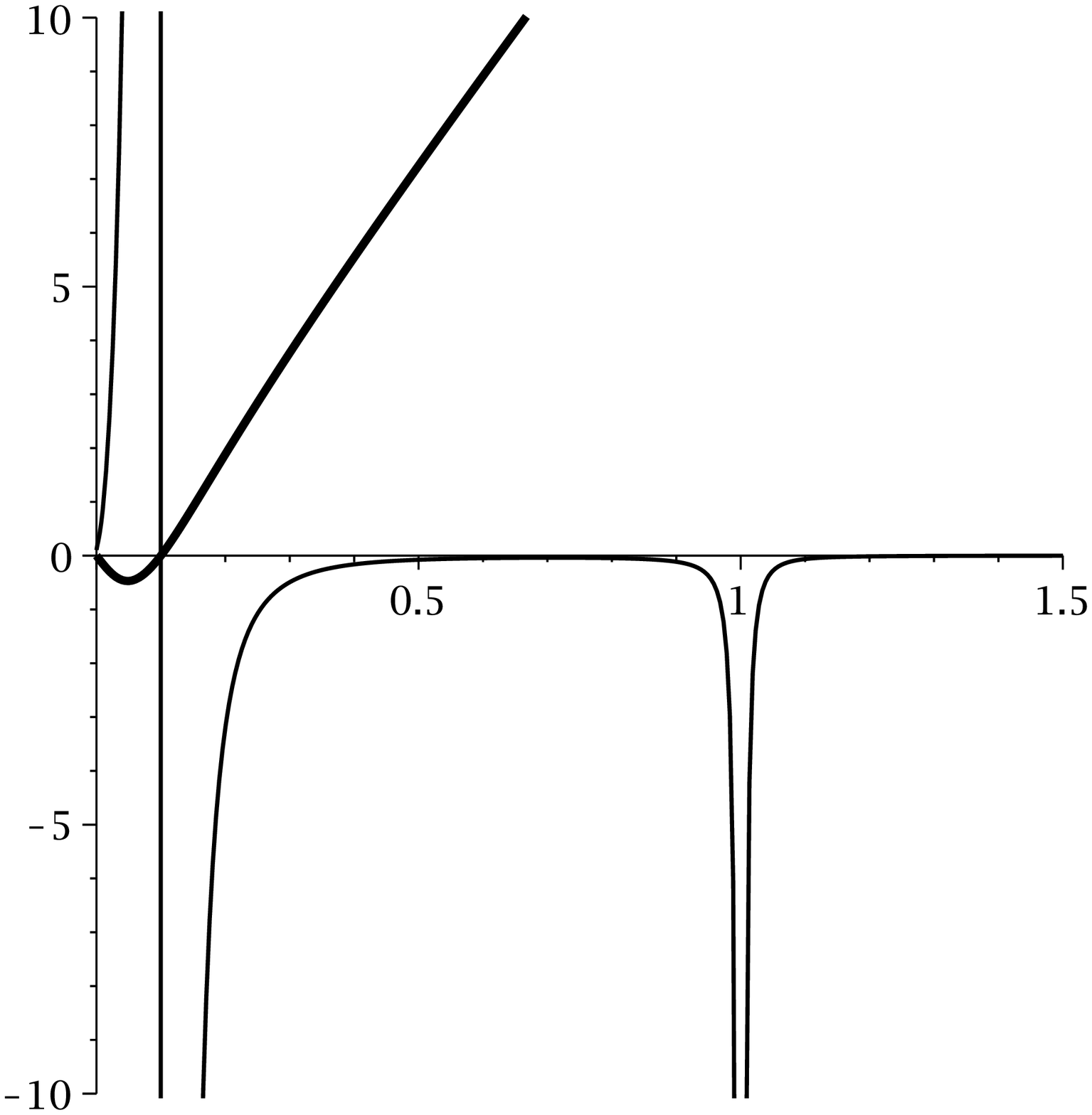}%
\end{array}
$%
\caption{\textbf{Maxwell case:} $\mathcal{R}$ and $C_{Q}$ versus
$r_{+}$ for $l=1$, $\Lambda =-1$, $d=3$ and $q=0.1$. \newline TRS
(continuous line) and heat capacity (bold line) for the Weinhold
metric (left-up panel), the Ruppeiner metric (right-up panel), the
Quevedo metric for case I (left-down panel) and the Quevedo metric
for case II (right-down panel). } \label{Fig6}
\end{figure}

\begin{figure}[tbp]
$%
\begin{array}{ccc}
\epsfxsize=5.5cm \epsffile{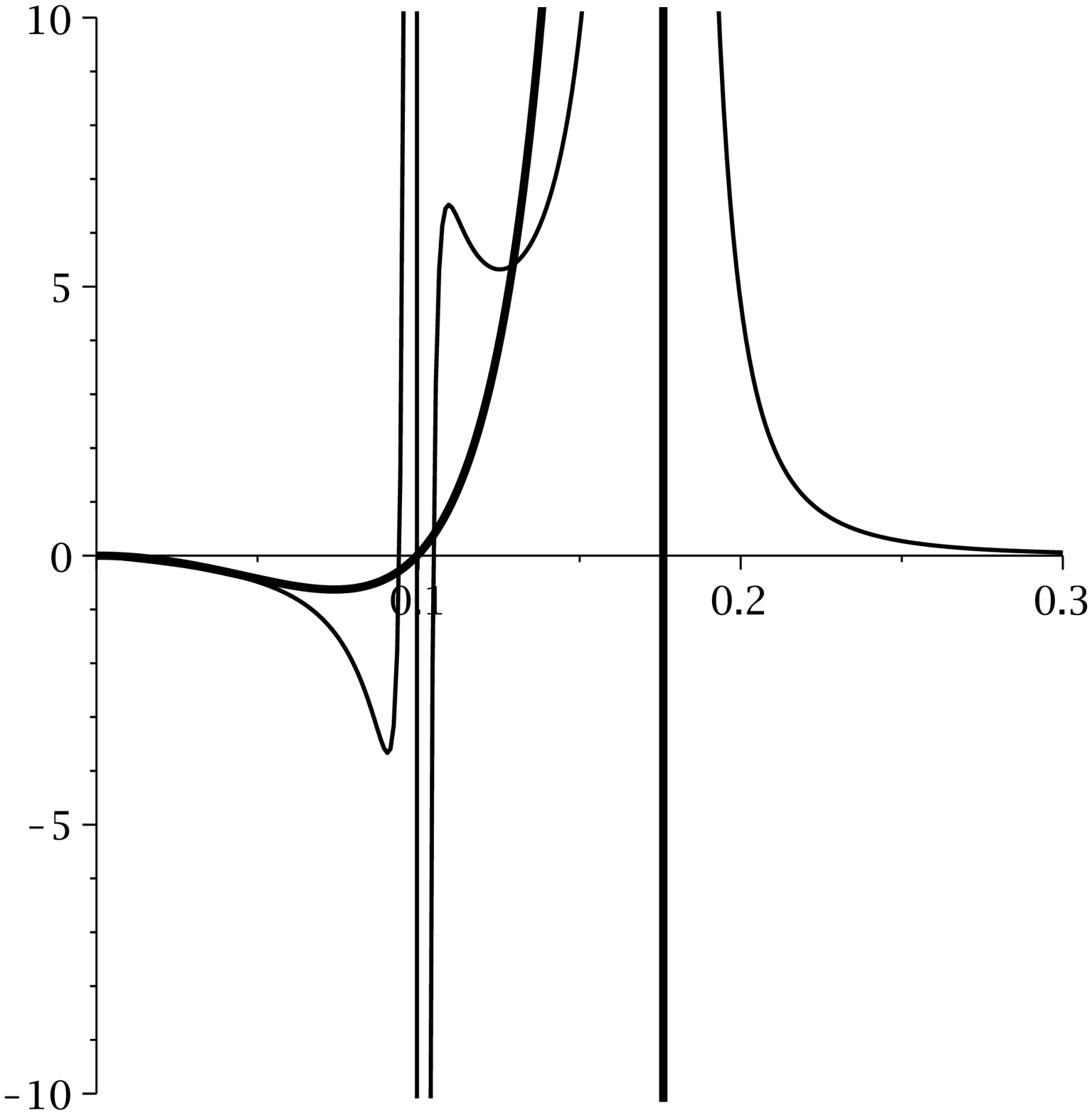} & \epsfxsize=5.5cm %
\epsffile{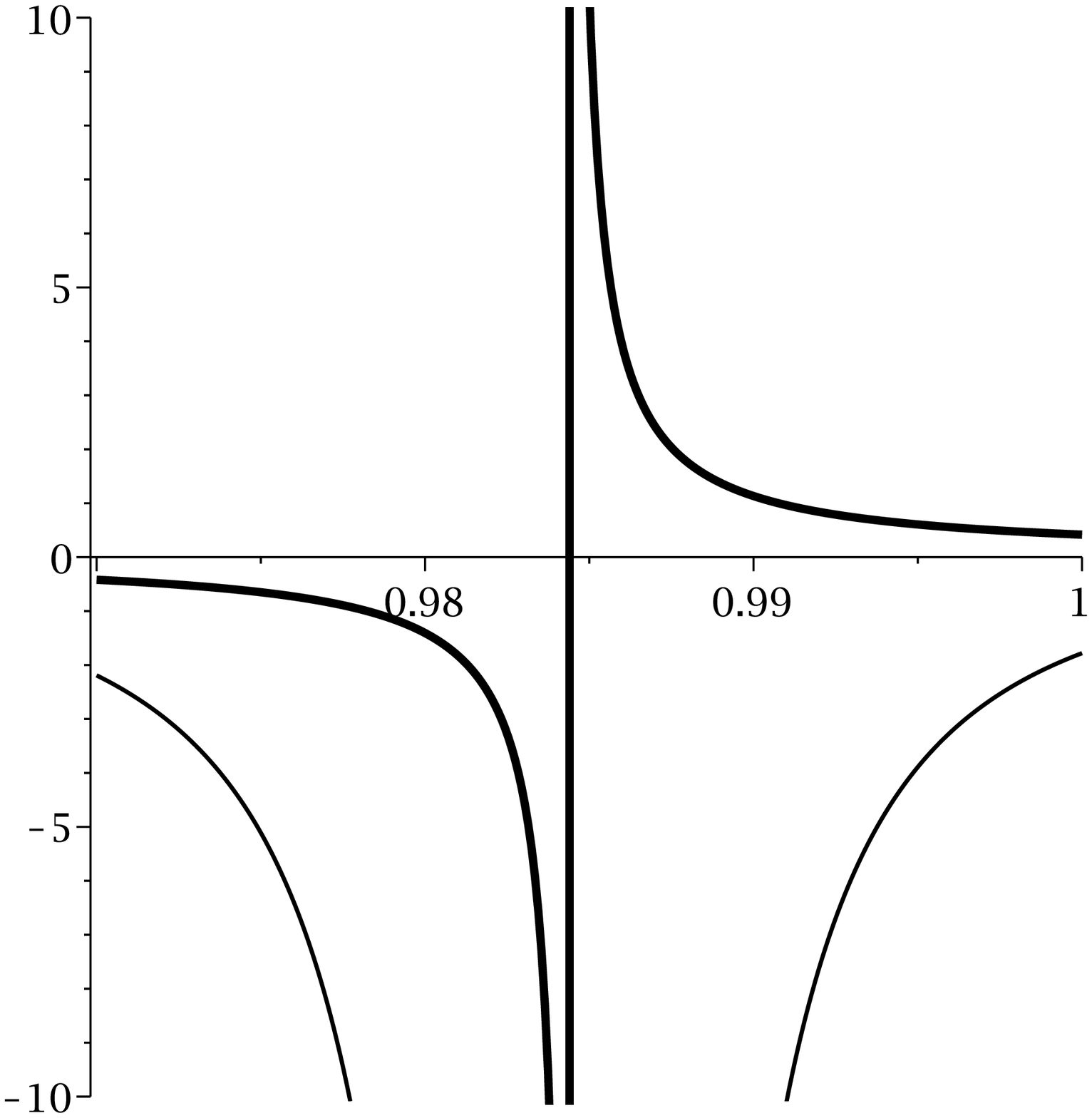} & \epsfxsize=5.5cm \epsffile{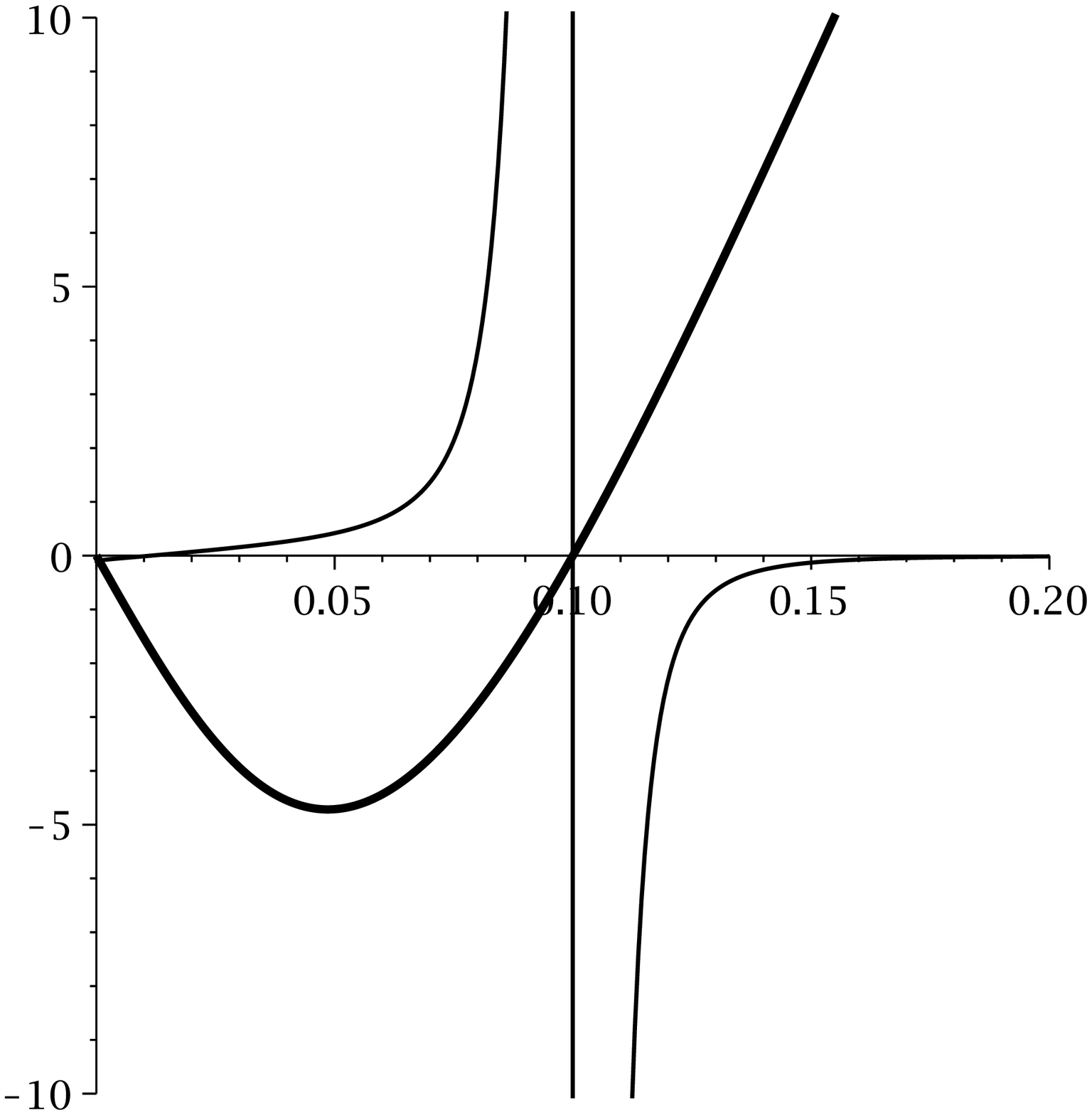}%
\end{array}
$%
\caption{\textbf{Maxwell case:} $\mathcal{R}$ and $C_{Q}$ versus
$r_{+}$ for $l=1$, $\Lambda =-1$, $q=0.1$. \newline TRS
(continuous line) and heat capacity (bold line) for the new metric
with $d=4$ (left and middle panels for different scales) and $d=3$
(right panel)} \label{Fig7}
\end{figure}

Using Eq. (\ref{massMax}), we are in a position to study the
behavior of TRS for different methods that were mentioned in this
paper. We can use Eqs. (\ref{Wein}), (\ref{Rupp}), (\ref{Quev})
and (\ref{newmetric}) with the heat capacity relations of
Einstein-Maxwell solutions, Eq. (\ref{HC}), to plot various
figures for studying the geometrical behavior of different
thermodynamical spacetime (see Figs. \ref{Fig4} - \ref{Fig7}).

Regarding Figs. \ref{Fig4} - \ref{Fig7}, we find that for $3$-dimensional
Einstein-Maxwell solutions (with considered values for different
parameters), there is only one root for the heat capacity. In other words,
in this case, we see one phase transition of type one and there is no
divergency for the heat capacity. Whereas in $4$-dimensional solutions,
there are one phase transition of type one and two phase transition of type
two. It means that there are three phase transition points in four
dimensional Einstein-Maxwell black hole solutions.

Regarding Weinhold approach, one finds both mismatching and extra
divergencies in three and four dimensions (see Figs. %
\ref{Fig4} and \ref{Fig6}). In other words, in $4$-dimensional black holes
(Fig. \ref{Fig4}), the root and one of the divergencies of the heat capacity
are not compatible with any divergency of the Weinhold's TRS. On the other
hand, in $3$-dimensional solutions (Fig. \ref{Fig6}), this approach
completely fails to provide an effective machinery for studying the phase
transitions. One divergency is observed which does not coincide with any
phase transition point and for the root of heat capacity TRS is a smooth
function.

As for the Ruppeiner metric, in $4$-dimensional Einstein-Maxwell solutions
(Fig. \ref{Fig4}), one of the divergencies of the heat capacity is not
matched with any divergency of the Ruppeiner's TRS. For $3$-dimensional
case, there are two divergencies for TRS (Fig. \ref{Fig6}). One of them
coincides with the root of heat capacity whereas the other one is not
related to any phase transition point.

Now we investigate both types of Quevedo metrics and their
behaviors. Regarding case I in $4$-dimensional solutions (Fig.
\ref{Fig5} up panels), there is no divergency for TRS in the place
of heat capacity vanishing point, whereas divergencies of the heat
capacity are matched with divergencies of TRS. For case II (Fig.
\ref{Fig5} down panels), we obtain effective results. In other
words, all divergence points and vanishing point of the heat
capacity are matched with divergencies of the Quevedo's TRS of
case II. But for charged BTZ black holes, regardless of case I or
II, an extra divergency is seen which is not related to any phase
transition point of the heat capacity (Fig. \ref{Fig6} down
panels).

Finally, for both $3$ and $4$ dimensional Einstein-Maxwell black
holes (see Fig. \ref{Fig7}), new metric is completely successful
for describing phase transition points of the heat capacity in
context of geometrothermodynamics. In other words, regardless of
the dimensions, the phase transition points of the heat capacity
and divergencies of the Ricci scalar of the constructed spacetime,
coincide and this approach is free of extra divergencies for its
TRS.

We should note that this new metric enjoys the characteristic behavior which
was mentioned in Einstein-BI solutions and one can distinguish two types of
phase transitions from each other.

\subsection{Neutral (uncharged) solutions}

In this section, we study the behavior of the system in case of
uncharged solutions ($q=0$). Considering the uncharged solutions
with an extensive parameter (entropy), one can find that the TRS
of mentioned methods of geometrothermodynamics vanishes.
Therefore, it seems that it is not possible to use these methods
for studying the behavior of phase transitions. In other words,
the geometrothermodynamics methods are valid for systems
containing two or more extensive parameters.

\section{Generalization}

Now, we are in a position to generalize the new defined metric to
the case of more than two extensive parameters. Regarding
Weinhold, Ruppeiner and Quevedo approaches, one finds that
supplementing additional extensive parameters increases the
complexity of the thermodynamical Ricci scalars and also their
denominators. The supplemental extensive parameters lead to extra
terms which may contribute to the additional number of
divergencies of the system under study. Considering the total mass
of the system as a function of arbitrary number of extensive
parameters $\chi _{i}$'s, one finds that the denominator of TRS
for Weinhold, Ruppeiner and Quevedo approaches will be more
complicated. Numerical calculations show that denominator of TRS
for the mentioned geometrical approaches contain extra terms which
may contribute to number of the TRS divergencies. For more
clarifications, in addition to $S$ and $Q$, we consider angular
momentum, $J$, as extensive parameter. Regarding the total mass of
the black holes as a function of $S$, $Q$ and $J$, we calculate
the denominator of TRS for the Weinhold, the Ruppeiner and the
Quevedo metrics
\begin{equation}
denom(\mathcal{R})=\left\{
\begin{array}{cc}
M^{3}\xi  & Weinhold\vspace{0.3cm} \\
M^{3}T^{3}\xi  & Ruppeiner\vspace{0.3cm} \\
M_{SS}^{2}M_{QQ}^{2}M_{JJ}^{2}\left( SM_{S}+QM_{Q}+JM_{J}\right) ^{3} &
Quevedo\;\;Case\;\;I\vspace{0.3cm} \\
S^{3}M_{SS}^{2}M_{QQ}^{2}M_{JJ}^{2}M_{S}^{3} & Quevedo\;\;Case\;\;II%
\end{array}%
\right. ,  \label{DenOthers}
\end{equation}%
where
\begin{equation}
\xi =\left[ M_{SS}\left( M_{QJ}^{2}-M_{QQ}M_{JJ}\right)
+M_{SQ}^{2}M_{JJ}+M_{SJ}^{2}M_{QQ}-2M_{SQ}M_{SJ}M_{QJ}\right] ^{2}.  \notag
\end{equation}

Considering Eq. (\ref{DenOthers}), one finds that although
vanishing points of heat capacity coincide with related divergence
points of TRS in the Ruppeiner case (due to the existence of $T$),
for the Weinhold one the coincidences are not generally observed.
Regarding the divergence points of the heat capacity, which are
related to vanishing $M_{SS}$, and in order to match both
divergence points of heat capacity and those of the Weinhold and
the Ruppeiner cases, one should set last three terms to zero and
the coefficient of $M_{SS}$ should be a nonzero finite expression
($M_{QJ}^{2}-M_{QQ}M_{JJ}\neq 0$). Since these fine tuning
conditions are not hold in general, we encounter mismatch between
divergence points and/or extra divergencies in TRS.

As for Quevedo metrics, due to existence of $M_{SS}^{2}$,
divergencies of both heat capacity and Ricci scalar match with
each other. As for the phase transition type one and in order to
have coincidence between roots of the heat capacity and divergence
points of the TRS, for the case $II$ Quevedo metric, existence of
$M_{S}^{3}$ is sufficient to ensure the mentioned matching whereas
we have the following restriction for the case $I$

\begin{equation}
SM_{S}+QM+JM_{J}=M_{S}.  \label{Res1}
\end{equation}

In general Eq. (\ref{Res1}) is not hold for arbitrary metric
function parameters and therefore, there will be extra
divergencies for the Ricci scalar which are not matched with any
phase transition point. Regarding both of Quevedo metrics,
$M_{QQ}^{2}$ and $M_{JJ}^{2}$, may contribute to
additional divergencies of the Ricci scalar which was seen in case of $3$%
-dimensional solutions.

In order to solve these problems we generalize the new introduced
thermodynamical metric to the case of $n$ extensive parameters ($n \geq 2$)
with following form
\begin{equation}
ds_{New}=\frac{SM_{S}}{\left( \Pi _{i=2}^{n}\frac{\partial ^{2}M}{\partial
\chi _{i}^{2}}\right) ^{3}}\left( -M_{SS}dS^{2}+\sum_{i=2}^{n}\left( \frac{%
\partial ^{2}M}{\partial \chi _{i}^{2}}\right) d\chi _{i}^{2}\right) ,
\label{GenMet}
\end{equation}
where $\chi _{i} \neq S$. In order to obtain the curvature
singularity of the new generalized metric, we should calculate the
Ricci scalar. Since analytical calculations are too large, for the
sake of brevity, we study the denominator of the Ricci scalar.
Calculations show that the too long expression of the numerator of
TRS (generalization of Eq. (\ref{NumNew})) is divergence free
\cite{Singular}, while its denominator is
\begin{equation}
denom(\mathcal{R})=S^{3}M_{S}^{3}M_{SS}^{2}.  \label{DenGen}
\end{equation}
Eq. (\ref{DenGen}) confirms that all singular points of TRS coincide with
both types of heat capacity phase transition points without any extra
divergency.


\section{Conclusions}

Motivated by a surge of study of geometrical concept of black hole
thermodynamics, we introduced a new metric regarding the matter.
Considering the various approaches toward geometrical study of
black hole thermodynamics, we first discussed the shortcomings of
the mentioned methods. It was believed that the divergencies of
TRS indicate the thermodynamical phase transitions, and therefore
we focused on the roots of denominator of TRS (since we regarded
smooth function of mass and its derivatives, the numerator of TRS
is a regular function and therefore, divergencies of TRS is
equivalent to the roots of denominator of TRS). Taking into
account the Weinhold, Ruppeiner and different types of Quevedo
metrics, we showed that divergencies of TRS may not coincide with
roots and divergencies of the corresponding heat capacity.

In order to avoid this problem, we introduced a new metric. We
showed that the denominator of its TRS contains terms which are
only the product of numerator and denominator of the corresponding
heat capacity. In other words, all divergencies of TRS in this
approach coincide with phase transition points of the heat
capacity. Moreover, for more clarifications, we regarded two known
examples to show the shortcomings of the other previous metrics
and the efficiency of the new devised metric.

Next, we generalized these metrics to contain more than two extensive
parameters. As it was seen, in the cases of Weinhold, Ruppeiner and
different types of Quevedo metrics, (denominator of) TRS contained
complicated expressions and extra terms that may increase the number of TRS
divergencies and shift the place of these divergencies in a way that they
may not coincide with phase transition points arisen from the heat capacity.
We showed that the divergencies of TRS related to the generalized new
introduced metric are compatible to the phase transition points of heat
capacity.

Another important property of the new metric is the different behavior of
TRS before and after its divergence points. It was seen that the behavior of
TRS for divergence points related to two types of the phase transition is
different. Therefore, considering this approach also enable us to
distinguish these two types of phase transition from one another.

Recently, it was seen and proposed that different constants (such
as cosmological constant, BI nonlinearity parameter, Gauss-Bonnet
parameter, Newton constant and etc.) may vary and have
contribution to thermodynamical structure of the system
\cite{vari}. In other words, in case of black holes, the total
mass of the black hole is a function of these parameters as
extensive parameters. It will be interesting to reconsider these
constants as thermodynamical variables and modify TRS of the
mentioned geometrothermodynamical methods. Although this
modification changes TRS of various methods, it does not cause
inconsistent results for new generalized metric.

The approach that we introduced in this paper enable one to map the
divergence points of its TRS with phase transition points without any
concern regarding contribution of other terms and extra divergence points.
This method may also be employed to study phase transition of other non
gravitating systems.

\begin{acknowledgements}
We gratefully thank the anonymous referee for enlightening
comments and suggestions which substantially helped in proving the
quality of the paper. We also wish to thank the Shiraz University
Research Council. This work has been supported financially by the
Research Institute for Astronomy and Astrophysics of Maragha,
Iran.
\end{acknowledgements}


\begin{thebibliography}{99}
\bibitem{Singular} To our knowledge various black holes in gravitational theories have smooth functions of finite
mass. Nevertheless, we exclude the possible black holes with
non-smooth finite mass and its derivatives.
\end{thebibliography}
\end{document}